\documentclass[aps,prresearch,twocolumn,amsmath,amssymb,superscriptaddress,longbibliography]{revtex4-2}

\usepackage{hyperref}

\usepackage{times}
\usepackage[utf8x]{inputenc}
\usepackage[english]{babel}
\usepackage[T1]{fontenc}
\usepackage{lmodern}
\usepackage{amssymb}
\usepackage{amsmath}
\usepackage{amsthm}
\usepackage[pdftex]{graphicx}
\usepackage{epstopdf}
\usepackage{braket}
\usepackage[bottom]{footmisc}
\usepackage{dcolumn}
\usepackage{bm}
\usepackage{color}
\usepackage{relsize,dsfont,mathrsfs,empheq,verbatim,upgreek}
\usepackage{etoolbox}
\usepackage[caption = false]{subfig}
\usepackage{enumerate}
\usepackage{float}
\usepackage{xcolor}

\newcommand{\mathd}{\mathrm{d}}
\newcommand{\Tr}{\mathrm{Tr}}
\newcommand{\Lt}{\mathcal{L}}

\renewcommand{\AA}{\mathcal{A}}

\newcommand{\CC}{\mathcal{C}}

\newcommand{\LL}{\mathcal{L}}

\newcommand{\ii}{\text{i}}
\newcommand{\tr}[2]{\text{Tr}_{ #1 } \left\{ #2 \right\}}

\newcommand{\tS}{\text{S}}

\newcommand{\tE}{\text{E}}
\newcommand{\tB}{\text{B}}

\newcommand{\e}{\text{e}}

\newcommand{\ph}{{\phantom{\dagger}}}
\renewcommand{\bm}[1]{\textbf{\textit{#1}}}

\newcommand{\ac}[2]{\textcolor{violet}{}}

\begin{document}


\title{Quantum Otto cycle in the Anderson impurity model}


\author{Salvatore Gatto}
\email{salvatore.gatto@physik.uni-freiburg.de}
\affiliation{Institute of Physics, University of Freiburg, Hermann-Herder-Strasse 3, D-79104 Freiburg, Germany}

\author{Alessandra Colla}
\email{alessandra.colla@physik.uni-freiburg.de}
\affiliation{Institute of Physics, University of Freiburg, Hermann-Herder-Strasse 3, D-79104 Freiburg, Germany}
\affiliation{Dipartimento di Fisica ``Aldo Pontremoli'', Università degli Studi di Milano, Via Celoria 16, I-20133 Milan, Italy}
\affiliation{INFN, Sezione di Milano, Via Celoria 16, I-20133 Milan, Italy}

\author{Heinz-Peter Breuer}
\affiliation{Institute of Physics, University of Freiburg, Hermann-Herder-Strasse 3, D-79104 Freiburg, Germany}
\affiliation{EUCOR Centre for Quantum Science and Quantum Computing, University of Freiburg, Hermann-Herder-Str. 3, D-79104 Freiburg, Germany}

\author{Michael Thoss}
\affiliation{Institute of Physics, University of Freiburg, Hermann-Herder-Strasse 3, D-79104 Freiburg, Germany}

\begin{abstract}

We study the thermodynamic performance of a periodic quantum Otto cycle operating on the single-impurity Anderson model.
Using a decomposition of the time-evolution generator based on the principle of minimal dissipation, combined with the numerically exact hierarchical equations of motion (HEOM) method, we analyze the operating regimes of the quantum thermal machine and investigate effects of Coulomb interactions, strong system–reservoir coupling, and energy level alignments. 
Our results show that Coulomb interaction can change the operating regimes and may lead to an enhancement of the efficiency.

\end{abstract}

\date{\today}

\maketitle

\section{Introduction}\label{sec:Intro}

Quantum thermodynamics has evolved into a vibrant area of research over the past years, driven by both foundational questions and experimental progress \cite{Gemmer2004,Schaller2014,Binder2018,Deffner2019,Landi2021}. In particular, quantum thermal machines have gathered increasing attention, with several experimental implementations already demonstrating their feasibility \cite{Abah2012,Rossnagel2016,deAssis2019,Lindenfels2019,Peterson2019,Myers2022,Kosloff2015,Alicki2015,Liu2021,Cangemi2024,Koyanagi2024}.
The thermodynamic performance of quantum dots has also been widely studied in autonomous, steady-state setups without time-dependent driving \cite{Esposito2009,Kennes2013,Josefsson2019,Pyurbeeva2025,Volosheniuk2025}.

While the framework of quantum thermodynamics is well established in the weak-coupling regime of open quantum systems, many questions remain when going beyond this limit \cite{Breuer2016a,deVega2017,Campell2018,Strasberg2019,Rivas2020,Alipour2017}. In parameter regimes of stronger system-environment coupling, conventional assumptions break down and the definition of key thermodynamic quantities -- such as work, heat, and entropy production -- require careful analysis.
A central challenge lies in the intricate interplay between a system and its environment. As the coupling strength increases, the interaction energy between system and environment becomes significant. It is still debated to what extent this energy can be attributed to the system itself and how it influences the system's thermodynamic behavior \cite{Picatoste2025,Colla2025thermal}.

Numerous theoretical approaches have been proposed to address these issues, yet a full understanding remains elusive \cite{Weimer2008,Esposito2010,Teifel2011,Alipour2016,Seifert2016,Strasberg2017,Rivas2020,Alipour2021,Landi2021}. Recent studies have explored how strong coupling and memory effects impact the performance of quantum engines \cite{Segal2021,Zhang2014,Pozas-Kerstjens2018,Thomas2018,Pezzutto2019,Mukherjee2020,Wiedmann2020,Wiedmann2021,Chakraborty2022,Kaneyasu2023,Ishizaki2023,Liu2021,Gelbwaser2015}. A promising strategy to address these questions involves a revised formulation of thermodynamic laws, wherein the principle of minimal dissipation is used to uniquely decompose the generator of the system dynamics into a dissipative term and a Hamiltonian contribution, leading to the definition of an effective Hamiltonian $K_{\rm S}$ \cite{Colla2022a, Picatoste2023, Gatto2024}.

In this work, we apply this framework to analyze a quantum Otto cycle where a central interacting system is alternately coupled to fermionic environments held at different temperature. The working medium is described by the single-impurity Anderson model, consisting of a single electronic level that can be doubly occupied by electrons with opposite spin which interact via on-site Coulomb repulsion. The time evolution is simulated using the numerically exact hierarchical equations of motion (HEOM) method \cite{Tanimura1989,Thoss2021}, which is particularly suitable to efficiently recover open system dynamics and the effective Hamiltonian $K_S$, allowing us to access thermodynamic quantities very accurately.

Considering a broad range of parameters, we analyze the operating regimes of the quantum thermal machine. In particular, we show how the Coulomb interaction influences the prevailing operting regime and the efficiency. Notably, we find that while Coulomb interaction tends to hinder efficiency when energy levels are above the Fermi energy, they can actually enhance performance in the opposite regime. This intriguing result is also found when other definitions of energy and work are used and is thus a robust finding. We also explore how the effective Hamiltonian $K_{\rm S}$ deviates from the bare system Hamiltonian $H_{\rm S}$ due to the interaction with the leads, and how this influences the thermodynamic behavior of the cycle. 
Furthermore, we investigate how the effective Hamiltonian $K_{\rm S}$ accounts for the system's internal energy, including contributions from the interaction with the environment, by comparing $\langle K_{\rm S} \rangle$ to perturbative estimates based on the system-environment coupling strength.

The paper is organized as follows. In Sec.~\ref{sec:qthermo}, we recapitulate the principle of minimal dissipation and introduce the relevant thermodynamic quantities. The single-impurity Anderson model is presented in Sec.~\ref{sec:model}, while in Sec.~\ref{sec:heom} we introduce the HEOM approach for simulating the system's dynamics. In Sec.~\ref{sec:cycle}, we discuss the periodic Otto cycle and its application to the system. The results are presented in Sec.~\ref{sec:results}. We first analyze how the cycle efficiency depends on the energy level alignment and the strength of Coulomb interaction (Sec.~\ref{effeps}), and then examine two distinct regimes of the single-impurity Anderson model in detail, where the energy levels are above the Fermi level (Sec.~\ref{sec:above}) and below the Fermi level (Sec.~\ref{sec:below}), respectively. Finally, Sec.~\ref{sec:conclu} concludes with a summary.

\section{Theory}\label{sec:theory}

\subsection{Principle of minimal dissipation}\label{sec:qthermo}

To study non-equilibrium quantum thermodynamics, we consider an open quantum system $ \rm S$ interacting with an external environment $\rm E$. The total Hamiltonian governing the joint dynamics is given by
\begin{equation} \label{ham-total}
	H(t) = H_{\rm S}(t) + H_{\rm int}(t) + H_{\rm E} ,
\end{equation}
where $H_{\rm S}(t)$ and $H_{\rm E}$ represent the system and environment Hamiltonians, respectively, while $H_{\rm int}(t)$ describes their interaction. Both $H_{\rm S}(t)$ and $H_{\rm int}(t)$ may carry explicit time dependence to account for driving protocols or temporally modulated couplings. The reduced states of system and environment are denoted by $\rho_{\rm S}$ and $\rho_{\rm E}$, while the full state of the combined system is described by the density operator $\rho_{\rm S+E}$. Under the common assumption of an initially uncorrelated state, $\rho_{\rm S+E}(0) = \rho_{\rm S}(0) \otimes \rho_{\rm E}(0)$, the system's evolution is described by a dynamical map,
\begin{equation} \label{map}
	\rho_{\rm S}(t) = \Phi_t\rho_{\rm S}(0) ,
\end{equation}
where $\rho_{\rm S}(t)$ is obtained by tracing out the environmental degrees of freedom from the full density matrix,
\begin{align}
	\rho_{\rm S}(t) = \tr{\tE}{\rho_{\rm S+E}(t)}.
\end{align}

In this work we adopt the framework developed in Ref.~\cite{Colla2022a}, which formulates the system dynamics via a time-convolutionless master equation for $\rho_{\rm S}(t)$,
\begin{equation} \label{tcl-meq}
	\frac{d}{dt}\rho_{\rm S}(t) = \Lt_t\rho_{\rm S}(t) = -i \left[K_{\rm S}(t),\rho_{\rm S}(t)\right] + {\mathcal{D}}_t  \rho_{\rm S}(t).
\end{equation}
This equation features a time-local generator $\Lt_t$ that captures memory effects typical of non-Markovian dynamics \cite{Shibata1977,Shibata1979,Breuer2002}. The generator naturally separates into two contributions: a Hamiltonian part described by the commutator with a Hermitian operator $K_{\rm S}(t)$, and a dissipative term represented by
\begin{equation} \label{dissipator}
\hspace{-2mm}	{\mathcal{D}}_t \rho_{\rm S} = \sum_{\rm k}\theta_{\rm k}(t)\Big[L_{\rm k}(t)\rho_{\rm S} L_{\rm k}^{\dag}(t) - \frac{1}{2}\big\{L_{\rm k}^{\dag}(t)L_{\rm k}(t),\rho_{\rm S}\big\}\Big],
\end{equation}
where the real coefficients $\theta_{\rm k}(t)$ are time-dependent rates and $L_{\rm k}(t)$ are the Lindblad-type operators, which may also vary in time.

It is important to note that the decomposition of $\Lt_t$ into its Hamiltonian and dissipative components is not unique. However, by invoking the principle of minimal dissipation \cite{Colla2022a}, one can select a unique effective Hamiltonian $K_{\rm S}(t)$ such that the dissipator is minimized according to a specific operator norm. Following the approach in Ref.~\cite{Hayden2021}, this minimum is achieved when the Lindblad operators are traceless, thereby uniquely identifying $K_{\rm S}(t)$ as the internal Hamiltonian of the system.
This Hamiltonian represents a renormalization of the system energy levels, and corresponds to the Lamb-shift in the weak coupling regime. For more insight into how the system-bath coupling affects this renormalization, see \cite{Colla2025Ks}.

With this definition, the internal energy of the system is given by the expectation value
\begin{equation} \label{internal-energy}
	U_{\rm S} (t) = \Tr \{K_{\rm S}(t)\rho_{\rm S}(t)\}.
\end{equation}
The corresponding energy balance is then expressed through the first law of thermodynamics,
\begin{equation}\label{first-law}
	\Delta U_{\rm S} (t) \equiv U_{\rm S}(t) - U_{\rm S}(t_0) = W_{\rm S}(t) + Q_{\rm S}(t),
\end{equation}
where the work $W_{\rm S}(t)$ and heat $Q_{\rm S}(t)$ contributions are defined as
\begin{align}
	W_{\rm S}(t) &= \int_{t_0}^t d\tau \, \Tr \big\{ \dot{K}_{\rm S}(\tau) \rho_{\rm S}(\tau) \big\},  \label{work} \\
	Q_{\rm S}(t) &= \int_{t_0}^t d\tau \, \Tr \big\{ K_{\rm S}(\tau) \dot{\rho}_{\rm S}(\tau)  \big\}.  \label{heat}
\end{align}

In the weak-coupling limit, the effective Hamiltonian $K_{\rm S}$ approaches the bare system Hamiltonian $H_{\rm S}$, and the standard definitions are recovered \cite{Spohn1978,Lebowitz1978,Alicki1979,Kosloff1984,Kosloff2013,Deffner2008Feb,Alicki2018,Picatoste2025,Colla2025thermal},
\begin{align}
	W_{\rm w}(t) &= \int_{t_0}^t d\tau \, \Tr \big\{ \dot{H}_{\rm S}(\tau) \rho_{\rm S}(\tau) \big\},  \label{workweak} \\
	Q_{\rm w}(t) &= \int_{t_0}^t d\tau \, \Tr \big\{ H_{\rm S}(\tau) \dot{\rho}_{\rm S}(\tau)  \big\}. \label{heatweak}
\end{align}

Another relevant thermodynamic quantity is the total energy transfer from the environment, expressed as
\begin{align}
	Q_{\rm E}(t) = \int_0^t \mathd \tau \, \Tr_{\rm S+E} \big\{ H_{\rm E} \, 
	\dot{\rho}_{\rm S+E}(\tau) \big\},
\end{align}
which is often used as a heat-like contribution in situations beyond the weak-coupling regime \cite{Landi2021}, and in particular definitions of entropy production \cite{Esposito2010,Strasberg2017}. It can also be employed in an alternative definition for the efficiency of the cycle (see, e.g. \cite{Liu2021}, and Sec.~\ref{sec:cycle}).

\subsection{Single-impurity Anderson model}\label{sec:model}

The goal of this work is to investigate a quantum periodic Otto cycle in regimes of strong system-environment and intrasystem coupling taking full account of non-Markovian behavior and memory effects.  To this end, we consider a single-impurity Anderson model \cite{Anderson1961} described by the Hamiltonian
\begin{align}
	{H} = {H}_\tS + {H}_{\rm int} + {H}_\tE \, .
	\label{eq:general_Hamiltonian}
\end{align}
The system Hamiltonian ${H}_\tS$ describes a single level which can be occupied by two electrons with different spin,
\begin{align}
	{H}_\tS = \varepsilon (t) \left( d_{\uparrow}^{\smash{\dagger}} d_{\uparrow}
	+d_{\downarrow}^{\smash{\dagger}} d_{\downarrow}\right) + U d_{\uparrow}^{\smash{\dagger}} d_{\uparrow}
	d_{\downarrow}^{\smash{\dagger}} d_{\downarrow} \, ,
	\label{eq:Hamiltonian:system}
\end{align}
where $d_{\sigma}^\dagger$ ($d_{\sigma}$) creates (annihilates) an electron with spin $\sigma=\uparrow,\downarrow$ in the system level.
The Hamiltonian is characterized by a time-dependent energy $\varepsilon(t)$ and Coulomb interaction $U$ between the two spin states.
 The system Hilbert space is spanned by the four local states 
\(\{ \ket{0}, \ket{\downarrow}, \ket{\uparrow}, \ket{\downarrow\uparrow} \}\), 
corresponding respectively to the empty state, the singly occupied states with spin down and spin up, and the doubly occupied state.  
The populations of these states are given by the diagonal elements of the reduced density operator \(\rho_\tS\),
\begin{align}
	\rho_{00} &= \bra{0} \rho_\tS \ket{0}, \quad
	\rho_{\downarrow} = \bra{\downarrow} \rho_\tS \ket{\downarrow}, \\
	\rho_{\uparrow} &= \bra{\uparrow} \rho_\tS \ket{\uparrow}, \quad
	\rho_{\downarrow\uparrow} = \bra{\downarrow\uparrow} \rho_\tS \ket{\downarrow\uparrow}.
\end{align}

The environment consists of electrons described by two fermionic baths held at different temperatures,
\begin{align}
	H_\tE = \sum_{\substack{\rm k\in\mathcal{K}_h\cup\mathcal{K}_c \\ \sigma=\uparrow,\downarrow}}
	\varepsilon_{\rm k,\sigma}\, c^\dagger_{\rm k,\sigma} c_{\rm k,\sigma}
	= H_h + H_c \, ,
	\label{eq:Hamiltonian:fermionic_environment}
\end{align}
where $\mathcal{K}_{\rm h}$ and $\mathcal{K}_{\rm c}$ denote the sets of electronic environmental degrees of freedom belonging to the hot and cold reservoirs, respectively, and $\sigma\in\{\uparrow,\downarrow\}$ labels the spin degree of freedom.
Here, $c^\dagger_{\rm k,\sigma}$ ($c_{\rm k,\sigma}$) creates (annihilates) an electron with spin $\sigma$ and energy $\varepsilon_{\rm k,\sigma}$ in bath state $\rm k$.

Each bath interacts with the system through system-environment coupling terms, which are linear in the environment creation/annihilation operators,
\begin{align}
	{H}_{\rm int}  =\sum\limits_{\substack{\rm k\in\mathcal{K}_h\cup\mathcal{K}_c \\ \sigma=\uparrow,\downarrow}} g_{\rm k,\sigma}\left( d^\dag_{\sigma}c_{\rm k,\sigma} +d_{\sigma}c^\dag_{\rm k,\sigma} \right) \, .
	\label{eq:Hamiltonian:fermionic_coupling}
\end{align}
The influence of each fermionic environment on the system dynamics is encoded in its spectral density function,
\begin{align}
	J_{\rm K}(\varepsilon) 
	= 2\pi \sum_{\substack{\rm k \in \mathcal{K}_{\rm K} \\ \sigma=\uparrow,\downarrow}} 
	|g_{\rm k,\sigma}|^2 \, \delta(\varepsilon - \varepsilon_{\rm k,\sigma}) \, ,
	\label{eq:fermionic_Spectral_density_general_definition}
\end{align}
where $\mathcal{K}_{\rm K}$ denotes the set of bath degrees of freedom associated with environment $K\in\{\rm h,c\}$. 
Throughout this paper, we consider a spectral density of Lorentzian form,
\begin{align}
	J_{\rm K}(\varepsilon)
	= \Gamma \, \frac{W^2}{W^2 + (\varepsilon - \mu_{\rm K})^2} \, ,
	\label{eq:lorentzian_Spectral_density}
\end{align}
where $\Gamma$ denotes the overall coupling strength, $W$ is the characteristic bandwidth, and $\mu_{\rm K}$ is the chemical potential of environment $\rm K$.

The model considered here extends previous studies of quantum thermodynamics in impurity models by  Segal et al. \cite{Segal2021}. In contrast to the resonant level model studied in \cite{Segal2021}, the Anderson impurity model includes intrasystem Coulomb interaction giving rise to a rich physical behavior. Because it is an interacting quantum impurity model, it also requires advanced numerical methods to describe its dynamics and thermodynamics.

\subsection{HEOM method}\label{sec:heom}
To investigate the dynamics and thermodynamics of the single-impurity Anderson model, we employ the hierarchical equations of motion (HEOM) approach, which provides a numerically exact method for simulating open quantum systems \cite{Tanimura1989,Schinabeck2016,Thoss2021}. The central quantity of interest is the reduced density matrix $\rho_{\rm S}(t)$ describing the subsystem. We assume a factorized initial condition for the total state, $\rho_{\rm S+E}(0)=\rho_{\rm S}(0)\otimes\rho_{\rm E}(0)$, with the environment initially in a thermal state,
\begin{align} \label{gibbs-hb}
	\rho_{\rm E}(0)= \prod\limits_{\rm K=h,c} e^{- \beta_{\rm K} H_{\rm K} }/Z_{\rm K} .
\end{align} 
Here, $\beta_{\rm K}=1/(k_{\rm B}T_{\rm K})$, where $T_{\rm c}$ ($T_{\rm h}$) denotes the temperature of the cold (hot) bath.
Throughout this work, the chemical potentials $\mu_{\rm K}$ are set to zero, so that energies are measured with respect to the Fermi level, $\varepsilon_{\rm F}=0$.
Due to the Gaussian nature of the noninteracting reservoirs, one can derive a formally exact path-integral expression for $\rho_{\rm S}(t)$ using the Feynman-Vernon influence functional \cite{Feynman1951,Feynman1963,Tanimura1989,Jin2008}.
In this context, all relevant information about the coupling between system and environment is fully encoded in the environment correlation functions, which are directly determined by the environment spectral densities (cf. Eqs.~\eqref{eq:lorentzian_Spectral_density}). The correlation functions are given by
\begin{align}
	C_{\rm K}^{s} (t) = & \sum\limits_{\substack{\rm k \in \mathcal{K}_{\rm K}\\ \sigma=\uparrow,\downarrow}} |g_{\rm k,\sigma}|^2 \tr{\tB}{\left(c^s_{\rm k,\sigma}c^{\bar{s}}_{\rm k,\sigma} {\rm e}^{s i\varepsilon^\ph_{\rm k,\sigma} t}\right)\rho_{E}(0)} \nonumber\\
	=&\int_{-\infty}^\infty d\varepsilon\frac{J_{\rm K}(\varepsilon)}{2\pi} 
	\frac{{\rm e}^{s i\varepsilon t}}{{\rm e}^{s \beta_{\rm K} (\varepsilon-\mu_{\rm K})}+1} \, ,
\end{align}
with $c^{-(+)}_{\rm k,\sigma} \equiv c^{(\dagger)}_{\rm k,\sigma}$ and $s = \pm$. Within the HEOM method, these functions are approximated by sums over exponentials,
\begin{align}
	C_{\rm K}^s (t) = 
	\sum_{\rm l} \eta_{\rm l}^{s} \e^{-\gamma_{\rm l}^{s} t},
\end{align}
where the weights $\eta_{\rm l}^{s}$ and decay rates $\gamma_{\rm l}^{s}$ are computed using the Padé decomposition technique \cite{Hu2010,Hu2011,Kato2015,Erpenbeck2019,Schinabeck2020}.

The hierarchical equations of motion themselves are obtained through successive time derivatives of the influence functional, yielding \cite{Tanimura1989,Schinabeck2018},
\begin{alignat}{2}
	\frac{\partial}{\partial t} \rho^{(n)}_{\bm{j}}(t) =& -\left( i \LL_\tS + \sum_{r=1}^n \gamma_{j_r} \right) \rho^{(n)}_{\bm{j}}(t)
	\\
	&- \ii \sum_{j} \AA_j \rho^{(n+1)}_{\bm{j}j}(t) -\ii\sum_{r=1}^n (-)^{n-r} \CC_{j_r}  \rho^{(n-1)}_{\bm{j}\backslash j_r}(t)\nonumber ,
	\label{eq:general_HQME}
\end{alignat}
where $\bm{j} = \{j_1 \cdots j_n\}$ is a multi-index, $\bm{j}\backslash j_r$ denotes the multi-index with $j_r$ removed, and ${\bm{j}j} = \{j_1 \cdots j_n, j\}$. Each index $j$ corresponds to the set $\{K, l, m, \sigma\}$, and the superoperator $\LL_\tS$ acts as $\LL_\tS O = [H_\tS, O]$.

The lowest-tier element $\rho^{(0)}(t)$ is the reduced density matrix $\rho_{\rm S}(t)$, while higher-tier auxiliary density operators (ADOs) $\rho^{(n)}_{\bm{j}}(t)$ incorporate the non-Markovian influence of the environment. As $n$ increases, these ADOs capture increasingly higher-order system-environment correlations. Moreover, they encode environment-related observables such as energy and particle currents \cite{Jin2008,Haertle2013a,Kato2015,Kato2016}.

The operators $\AA_j$ and $\CC_j$ link ADOs at adjacent hierarchy levels via
\begin{subequations}  
	\begin{align}
		\AA_{j}\rho^{(n)}_{\bm{j}}=&  g_{\alpha,m}\left(d^{\sigma}_m\rho^{(n)}_{\bm{j}} + (-)^n \rho^{(n)}_{\bm{j}}d^{\sigma}_m \right) \, ,\\
		\CC_{j} \rho^{(n)}_{\bm{j}}=& g_{\alpha,m}\left(\eta_j d^{\sigma}_m\rho^{(n)}_{\bm{j}} - (-)^n \eta^*_j \rho^{(n)}_{\bm{j}}d^{\sigma}_m \right) \, ,
		\label{eq:general_HQME_upbuilding_operators}
	\end{align}
\end{subequations}  
resulting in a closed hierarchy of coupled differential equations.

\ac{}{(Maybe kill the following paragraphs in favour of the formula in \cite{Colla2025Ks}? It is more direct and skips pseudo-kraus, but gives the same result? In case, I can write something)}
In the following, we employ the HEOM method to compute the effective Hamiltonian $K_{\rm S}$ and evaluate thermodynamic quantities, such as work and heat, for the single-impurity Anderson model. 
To determine $K_{\rm S}$, we simulate the system dynamics for a set of orthonormal initial states $\ket{E_{\rm k}}$, allowing the reconstruction of the dynamical map $\Phi_t$ and its generator $\Lt_t$. Following the derivation in Ref.~\cite{Colla2025Ks}, the effective Hamiltonian is given by
\begin{align}
	K_{\rm S}(t) = -\frac{\ii}{4} \sum_{\rm j,k}  \Big[ \ket{E_{\rm k}}\bra{E_{\rm j}}, \Lt_t \left[ \ket{E_{\rm j}}\bra{E_{\rm k}}\right] \Big] \; ,
\end{align}
from which the thermodynamic observables are obtained via numerical integration.

\subsection{Periodic Otto cycle}\label{sec:cycle}

We study a quantum Otto cycle with period $\mathcal{T}$ using a single-impurity Anderson model with time dependent energy levels as working system. Our aim is to investigate the performance of the cycle for different parameter regimes, as well as comparing the thermodynamic observables in different formulations.
In the results presented below, $\mathcal{T}=30/\Gamma$ was used, which ensures that the system thermalizes within each stroke.

Throughout this work, a complete quantum Otto cycle \cite{Kosloff2017} with period $\mathcal{T}$ consists of four strokes (see Fig. \ref{ottocycle} for an illustration):  

\begin{figure}[t]
	\includegraphics[width=0.45\textwidth]{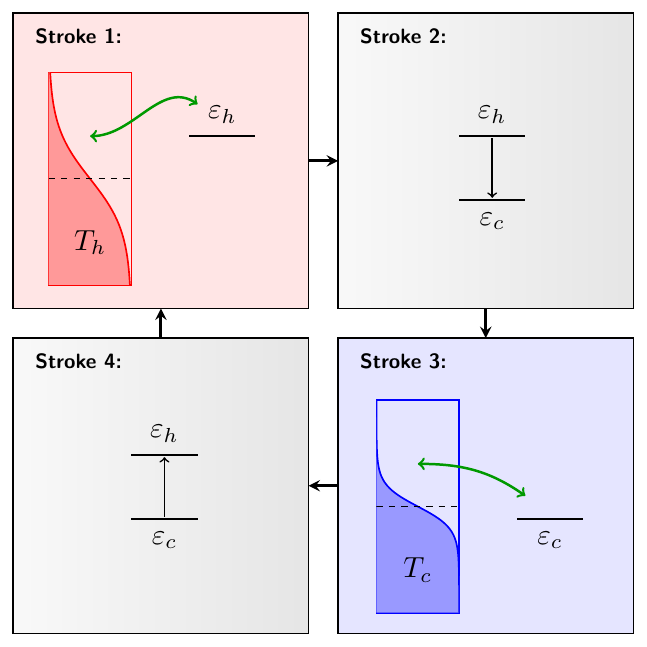} 
	\caption{Schematic of a periodic quantum Otto cycle.}
	\label{ottocycle}
\end{figure}

\begin{enumerate}[I.]
	
	\item Isochoric heating ($0\leq t<\mathcal{T}/3$): The system is connected to the hot bath at temperature $T_{\rm h}$, and evolves until thermalization is reached. During this stage, the energy levels are kept fixed at $\varepsilon(t)=\varepsilon_{\rm h}$. 	
	
	\item Adiabatic phase ($\mathcal{T}/3\leq t<\mathcal{T}/2$): The system is disconnected from the hot bath, and the energy levels of the system are driven from $\varepsilon_{\rm h}$ to $\varepsilon_{\rm c}$. 
	
	\item Isochoric cooling ($\mathcal{T}/2\leq t<5\mathcal{T}/6$): The system is connected to the cold bath at temperature $T_{\rm c}<T_{\rm h}$, and evolves with fixed energy levels $\varepsilon(t)=\varepsilon_{\rm c}$ until it thermalizes.
	
	\item Adiabatic phase ($5\mathcal{T}/6\leq t<\mathcal{T}$): The system is disconnected from the cold bath, and the energy levels of the system are driven back from $\varepsilon_{\rm c}$ to $\varepsilon_{\rm h}$. 
	
\end{enumerate}

In order to perform numerical simulations, it is necessary to specify the explicit time dependence of $\varepsilon(t)$. While the total work extracted over a full cycle does not depend on the specific shape of this protocol \cite{Liu2021}, it is nevertheless convenient to adopt a smooth parametrization. 
We therefore employ the following protocol over one period $[0,\mathcal{T}]$, where $\varepsilon(t)$ is constant during the isochoric strokes and varies smoothly during the adiabatic strokes:
\begin{equation}
	\varepsilon(t) =
	\begin{cases}
		\varepsilon_{\rm h} & 0\leq t<\mathcal{T}/3, \\
		\varepsilon_{\rm h} - (\varepsilon_{\rm h}-\varepsilon_{\rm c})\, Z(s_1) & \mathcal{T}/3\leq t<\mathcal{T}/2, \\
		\varepsilon_{\rm c} & \mathcal{T}/2\leq t<5\mathcal{T}/6, \\
		\varepsilon_{\rm c} + (\varepsilon_{\rm h}-\varepsilon_{\rm c})\, Z(s_2) & 5\mathcal{T}/6\leq t<\mathcal{T}.
	\end{cases}
	\label{eq:epsilon_t}
\end{equation}
Here $Z(s)=3s^2-2s^3$ ensures a smooth interpolation with vanishing derivatives at the boundaries, and the rescaled times are defined as $s_1=(t-t_1)/t_2$ and $s_2=(t-t_1-t_2-t_3)/t_4$. 

When working with this cycle, it is useful to define the heat contributions associated with strokes I and III, denoted respectively as $Q_{\rm h}$ and $Q_{\rm c}$,
\begin{align}
	Q_{\rm h} &= \int_{0}^{\mathcal{T}/3} d\tau \, \Tr \big\{ K_{\rm S}(\tau)\, \dot{\rho}_{\rm S}(\tau) \big\} , \nonumber \\
	Q_{\rm c} &= \int_{\mathcal{T}/2}^{5\mathcal{T}/6} d\tau \, \Tr \big\{ K_{\rm S}(\tau)\, \dot{\rho}_{\rm S}(\tau) \big\} ,  
	\label{heat_c}
\end{align}
while the total extracted work over one cycle is given by
\begin{align}
	W_{\rm tot} = \int_{0}^{\mathcal{T}} d\tau \, \Tr \big\{ \dot{K}_{\rm S}(\tau)\, \rho_{\rm S}(\tau) \big\} .
	\label{worktot}
\end{align}

We note that the switching on and off of the system--bath coupling at the boundaries between strokes may induce additional work-like contributions. These effects are effectively captured by discontinuities in the effective Hamiltonian $K_{\rm S}(t)$, which undergoes sudden quenches at those times \cite{Picatoste2023}. As a consequence, Eq.~\eqref{worktot} consistently accounts for all work contributions generated during the cycle. We also emphasize that system-environment correlations built up during the evolution are not reset at the end of each cycle.

Depending on the sign of $W_{\rm tot}$, $Q_{\rm h}$, and $Q_{\rm c}$, the system can operate in different thermodynamic modes:
\begin{itemize}
	\item \hspace{-0.6mm}$W_{\rm tot}<0$, $Q_{\rm h}>0$, $Q_{\rm c}<0$: heat engine
	\item \hspace{-0.6mm}$W_{\rm tot}>0$, $Q_{\rm h}<0$, $Q_{\rm c}>0$: refrigerator (heat pump)
	\item \hspace{-0.6mm}$W_{\rm tot}>0$, $Q_{\rm h}>0$, $Q_{\rm c}<0$: accelerator
	\item \hspace{-0.6mm}$W_{\rm tot}>0$, $Q_{\rm h}<0$, $Q_{\rm c}<0$: heater
\end{itemize}

Accordingly, we define the efficiency of the heat engine as
\begin{align}
	\eta_{\rm h} = -\frac{W_{\rm tot}}{Q_{\rm h}}.
	\label{def-etah}
\end{align}

In the analysis presented in the following sections, this strong-coupling efficiency will be compared with two other definitions of efficiencies commonly used in the literature.  
First, we consider a weak-coupling efficiency,
\begin{align}
	\eta_{\rm h}^{\rm w} = -\frac{W_{\rm tot}^{\rm w}}{Q_{\rm h}^{\rm w}},
	\label{def-etaw}
\end{align}
where $W_{\rm tot}^{\rm w}$ and $Q_{\rm h}^{\rm w}$ are defined analogously to Eqs.~\eqref{heat_c}-\eqref{worktot}, but using the bare system Hamiltonian $H_{\rm S}$ instead of the effective Hamiltonian $K_{\rm S}$.
Second, we consider the efficiency introduced by Esposito, Lindenberg and Van den Broeck \cite{Esposito2009},
\begin{align}
	\eta_{\rm h}^{\mathrm{ELB}} = \frac{W_{\rm tot}^{\rm w}}{Q_{\rm h}^{\rm E}},
	\label{def-etaelb}
\end{align}
where $W_{\rm tot}^{\rm w}$ is computed using $H_{\rm S}$, while $Q_{\rm h}^{\rm E}$ denotes the heat exchanged with the environment, evaluated through the environment Hamiltonian $H_{\rm E}$,
\begin{align}
	Q_{\rm h}^{\rm E} = \int_{0}^{\mathcal{T}/3} d\tau \, \Tr_{\rm S+E} \big\{ H_{\rm E}\, \dot{\rho}_{\rm S+E}(\tau) \big\} .   
	\label{heat_e}
\end{align}

\noindent
As a reference benchmark, we also consider the Carnot efficiency \cite{Kosloff2017},
\begin{align}
	\eta_{C} = 1 - \frac{T_{\rm c}}{T_{\rm h}},
	\label{def-etacarnot}
\end{align}
which sets an upper bound on the efficiency for an ideal reversible engine operating between the two temperatures $T_{\rm h}$ and $T_{\rm c}$.



\begin{figure*}[t!]
	\centering
	\includegraphics[width=\textwidth]{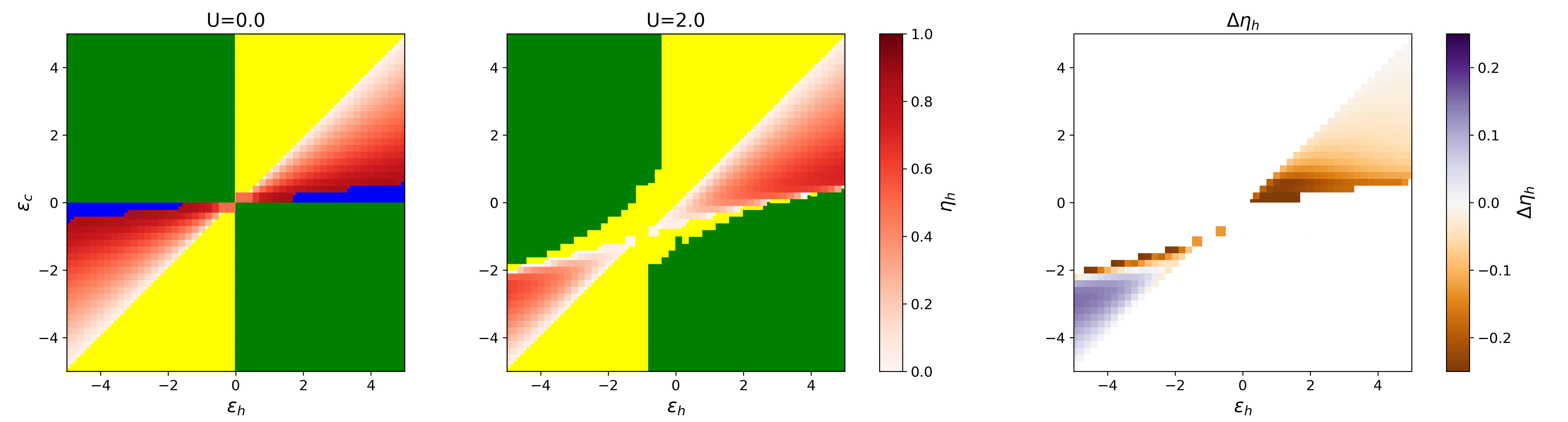}

	\caption{Left and center panels: Operating regimes of the quantum Otto cycle as a function of energy levels $\varepsilon_{\rm h}$ (horizontal axis) and $\varepsilon_{\rm c}$ (vertical axis) for Coulomb interactions $U=0$ (left) and $U=2$ (center). The colors  indicate different  regimes: heat engine (red), refrigerator (blue), accelerator (yellow), and heater (green).  The efficiency $\eta_{\rm h}$ is  shown for the heat engine regime. Right panel: Difference in efficiency for the heat engine regime, $\Delta \eta_{\rm h} = \eta_{\rm h}(U=2) - \eta_{\rm h}(U=0)$, indicating in which regimes the Coulomb interaction enhances the performance of the cycle.  Parameters, given in units of $\Gamma$, are $k_{\rm B} T_{\rm h} = 10$, $k_{\rm B} T_{\rm c} = 1$.}
	\label{fig:otto_cycle}
\end{figure*}

\section{Results}
\label{sec:results}
In this section, we use the concepts introduced above to analyze the quantum Otto cycle for a wide range of parameters. After an overview about the operating regimes and the efficiency of the cycle, we will focus on two distinct scenarios: energy levels above the Fermi level ($\varepsilon_{\rm h}, \varepsilon_{\rm c} > 0$), and energy levels below the Fermi level ($\varepsilon_{\rm h}, \varepsilon_{\rm c} < 0$).  For each scenario, we will investigate in detail the influence of the Coulomb interaction on the dynamics. 
Moreover, we analyze relevant thermodynamic quantities such as work $W_{\rm S}(t)$, heat $Q_{\rm S}(t)$, and the efficiency of the cycle as a heat engine, $\eta_{\rm h}$, which are then compared to the corresponding expressions defined in the weak-coupling limit, i.e., in the regime where the coupling strength $\Gamma$ is much smaller than the thermal energy scale ($\Gamma \ll k_{\rm B} T_{\rm h,c}$).

\subsection{Operating regimes and efficiency} \label{effeps}

We begin our analysis by examining the thermodynamic behavior of the quantum Otto cycle in terms of the energy levels $\varepsilon_{\rm h}$ and $\varepsilon_{\rm c}$ for the hot and cold configurations, respectively, and their interplay with the Coulomb interaction $U$. 
Fig.~\ref{fig:otto_cycle} presents a series of phase diagrams in the $(\varepsilon_{\rm h}  , \varepsilon_{\rm c}$) plane, where each point is color-coded according to the operational regime of the limit cycle. The classification is based on the sign of heat exchanged with the hot ($Q_{\rm h}$) and cold ($Q_{\rm c}$) reservoirs, as well as the net work $W_{tot}$, as explained in Section \ref{sec:cycle}. A detailed analytical derivation of the boundaries between these operational regimes in the weak coupling limit ($\Gamma \ll k_{\rm B} T_{h,c}$) is provided in Appendix~\ref{sec:weekcoupling}.

In the non-interacting case ($U=0$), the structure of the diagram is symmetric with respect to the origin, $(\varepsilon_{\rm h},\varepsilon_{\rm c})=(0,0)$, and the boundaries between the different regimes can be shifted by changing the parameters of the system. In the limit of weak system-environment coupling, the system operates as a heat engine (red) when both energy levels have the same sign and $  |\varepsilon_{\rm h}|T_{\rm c}/T_{\rm h} <|\varepsilon_{\rm c}| < |\varepsilon_{\rm h}|$ holds, absorbing heat from the hot reservoir and partially converting it into work, while dissipating the remaining energy into the cold reservoir. In the region with $  0 <|\varepsilon_{\rm c}| < |\varepsilon_{\rm h}|T_{\rm c}/T_{\rm h}$, the system behaves as a refrigerator (blue), a regime where external work is used to pump heat from the cold to the hot reservoir. This pattern of operating regimes is sensitive to the temperature ratio $T_{\rm c}/T_{\rm h}$: increasing $T_{\rm c}$ or decreasing $T_{\rm h}$ expands the refrigeration region. The accelerator regime (yellow) appears when $|\varepsilon_{\rm h}| < |\varepsilon_{\rm c}|$, provided that both energy levels lie on the same side of the Fermi level, either both above or both below. In this regime, the system absorbs heat from the hot reservoir and injects additional energy via external work, resulting in a faster transfer of heat to the cold reservoir than what would occur through passive conduction alone. Finally, when $\varepsilon_{\rm h}$ and $\varepsilon_{\rm c}$ lie on opposite sides of the Fermi level, the system operates in the heater regime (green), characterized by fully dissipative dynamics. In this regime, population inversion during the cycle is energetically unfavorable, and the external work performed on the system is entirely dissipated as heat into both reservoirs, without producing net transport or refrigeration. 

When Coulomb interaction is introduced ($U \ne 0$), the center of symmetry of the phase diagram moves to $(\varepsilon_{\rm h}, \varepsilon_{\rm c}) = (-U/2, -U/2)$ due to the energetic shift of the doubly occupied state (see below), and the phase boundaries are progressively deformed as $U$ increases. The most pronounced impact is observed in the refrigeration regime, which is pushed to higher values of $\varepsilon_{\rm h}$ and eventually disappears for large $U$, as heat absorption from the cold reservoir becomes energetically less favorable. This region is replaced by heater and accelerator behaviors, as the broadening of the doubly occupied energy level reduces the efficiency of converting heat into work, making the system increasingly dissipative.

The boundaries of the different regimes are also sensitive to the system–environment coupling strength $\Gamma$. In the weak-coupling limit, the effective Hamiltonian $K_{\rm S}$ closely matches the bare system Hamiltonian $H_{\rm S}$, and the corresponding thermodynamic observables, such as $Q_{\rm h}$, $Q_{\rm c}$, and $W_{\rm S}$, recover the analytical behavior discussed in Appendix~\ref{sec:weekcoupling} and illustrated in Fig.~\ref{fig:otto_working_regimes}. As the coupling strength is increased to moderate or strong values, $K_{\rm S}$ undergoes level shifts, corresponding to the renormalization of the energy levels due to the coupling with the environments \cite{Gatto2024, Colla2025natcomm}. 


The efficiency of the heat engine depicted in Fig.~\ref{fig:otto_cycle} is closely related to the boundaries of the operational regimes: it peaks near the boundary $\varepsilon_{\rm h} = \varepsilon_{\rm c} T_{\rm h}/T_{\rm c}$ between the engine and refrigerator regions, and decreases as one approaches $\varepsilon_{\rm h} = \varepsilon_{\rm c}$, where work extraction remains possible but less effective.
Introducing Coulomb interaction $U$ broadens the doubly occupied energy level, which in certain parameter regimes can effectively reduce the heat input required to achieve the same amount of work. This effect leads to an enhancement of the cycle efficiency compared to the noninteracting case. In particular, the efficiency is increased when the doubly occupied state tends to empty during the hot phase and refills during the cold phase (see Appendix~\ref{sec:weekcoupling}). Notably, this behavior is observed in the lower-left region of the phase diagram in Fig.~\ref{fig:otto_cycle}, where both energy levels lie below the Fermi level. The underlying mechanism originates from the asymmetry in the occupation of broadened energy levels induced by Coulomb interaction, providing a subtle route to boost efficiency and illustrating how quantum correlations can be leveraged to enhance thermodynamic performance.

\begin{figure}[t]
	\centering
	\begin{minipage}[b]{0.23\textwidth}
		\centering
		\includegraphics[width=\linewidth]{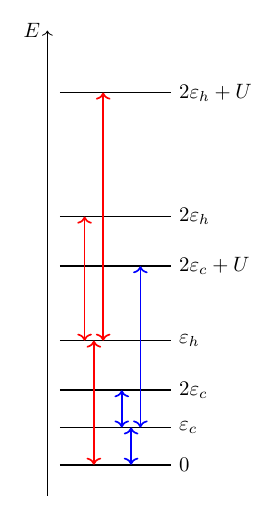} 
	\end{minipage}
	\hfill
	\begin{minipage}[b]{0.23\textwidth}
		\centering
		\includegraphics[width=\linewidth]{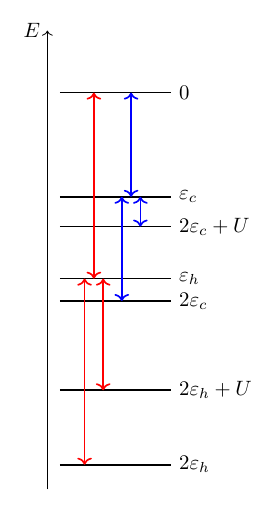} 
	\end{minipage}
	\caption{Energy-level schemes for the two configurations considered in Sec.~\ref{sec:above} and Sec.~\ref{sec:below}. 
		Left: levels above the Fermi energy. 
		Right: levels below the Fermi energy.}
	\label{fig:energy_levels}
\end{figure}

In the following two sections, we analyze the underlying mechanisms in some detail. Thereby, we concentrate on the heat engine regime and select two representative parameter regions from Fig.~\ref{fig:otto_cycle}, including one where both energy levels lie above the Fermi level ($\varepsilon_{\rm h}, \varepsilon_{\rm c} > 0$), and one where both levels lie below it ($\varepsilon_{\rm h}, \varepsilon_{\rm c} < 0$). The corresponding energy-level schemes are shown in Fig.~\ref{fig:energy_levels}. In each case, we consider the role of Coulomb interaction and its impact on the efficiency of the Otto cycle.

\subsection{Energy levels above the Fermi level}
\label{sec:above}

\begin{figure}[t]
	\includegraphics[width=0.50\textwidth]{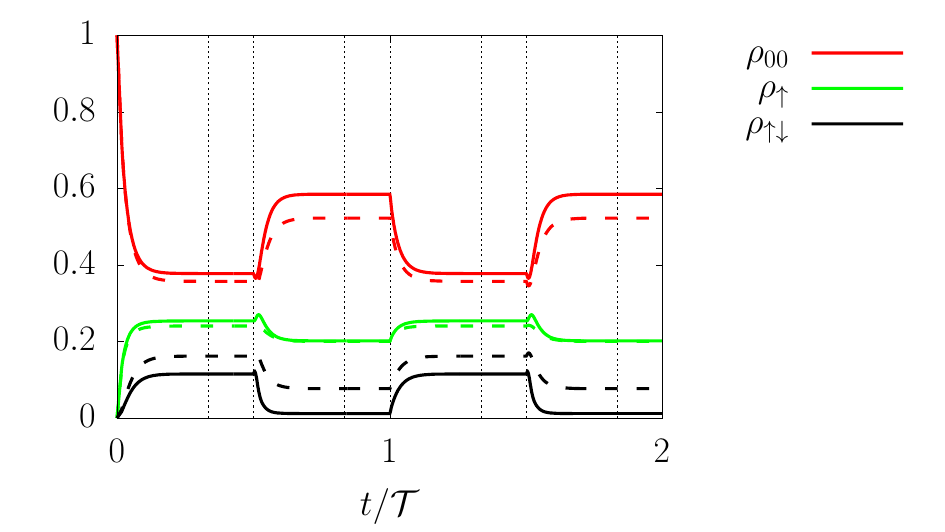} 
	\caption{Population of the different states of $\rho_{\rm S}(t)$ as a function of time. Dotted vertical lines indicate the boundaries between strokes of the Otto cycle. Parameters, given in units of $\Gamma$, are $U=0$ (dashed lines), $U=2$ (solid lines), $\varepsilon_{\rm h} =2$, $\varepsilon_{\rm c} =0.6$, $k_{\rm B}T_{\rm h}=10$, $k_{\rm B}T_{\rm c}=1$. }
	\label{popabove1}
\end{figure}
\begin{figure}[t]
	\includegraphics[width=0.50\textwidth]{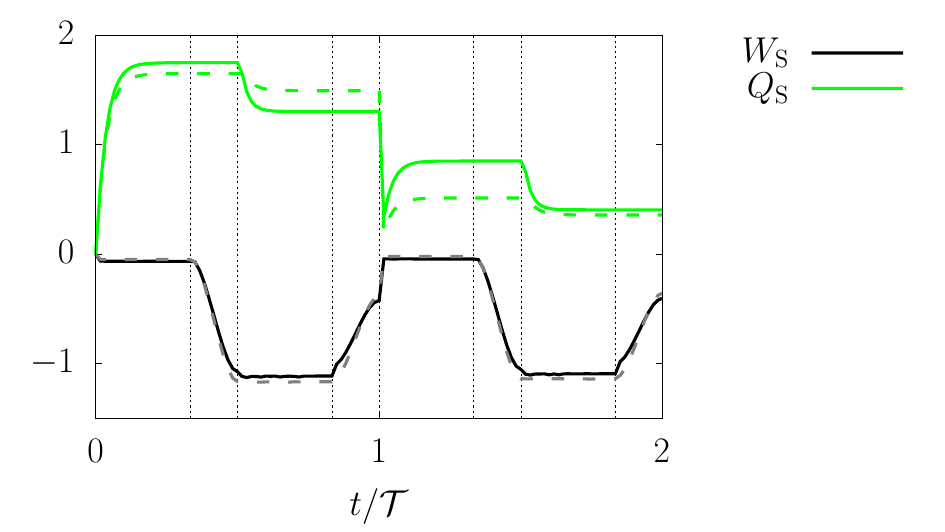} 
	\caption{Work $W_{\rm S}(t)$ and heat $Q_{\rm S}(t)$ as a function of time for the interacting  (U=2, solid lines) and the noninteracting  (U=0, dashed lines) case. Dotted vertical lines indicate the boundaries between strokes of the Otto cycle. Parameters are $\varepsilon_{\rm h} =2$, $\varepsilon_{\rm c} =0.6$, $k_{\rm B}T_{\rm h}=10$, $k_{\rm B}T_{\rm c}=1$. Parameters and results are given in units of $\Gamma$.}
	\label{termoabove1}
\end{figure}


\begin{figure}[t]
	\includegraphics[width=0.50\textwidth]{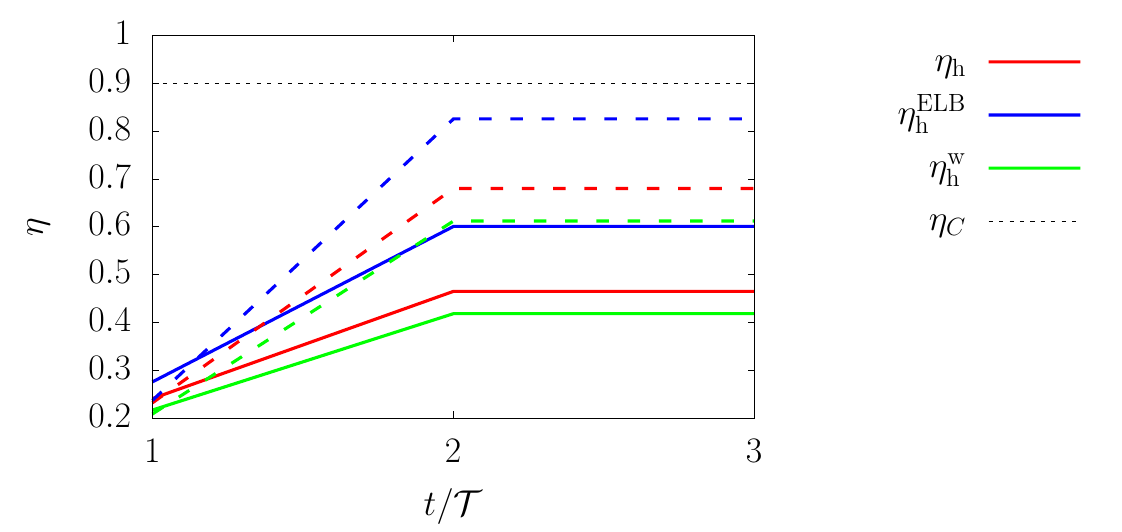} 
\caption{Comparison of the three efficiency definitions (Eqs.~\eqref{def-etah}--\eqref{def-etaelb}) 
	for the interacting ($U=2$, solid lines) and the noninteracting ($U=0$, dashed lines) case. 
	The black dotted line indicates the Carnot efficiency $\eta_C$. 
	Parameters, given in units of $\Gamma$, are $\varepsilon_{\rm h} = 2$, $\varepsilon_{\rm c} = 0.6$, 
	$k_{\rm B} T_{\rm h} = 10$, $k_{\rm B} T_{\rm c} = 1$.}
	\label{effabove1}
\end{figure}

\begin{figure}[t]
	\includegraphics[width=0.50\textwidth]{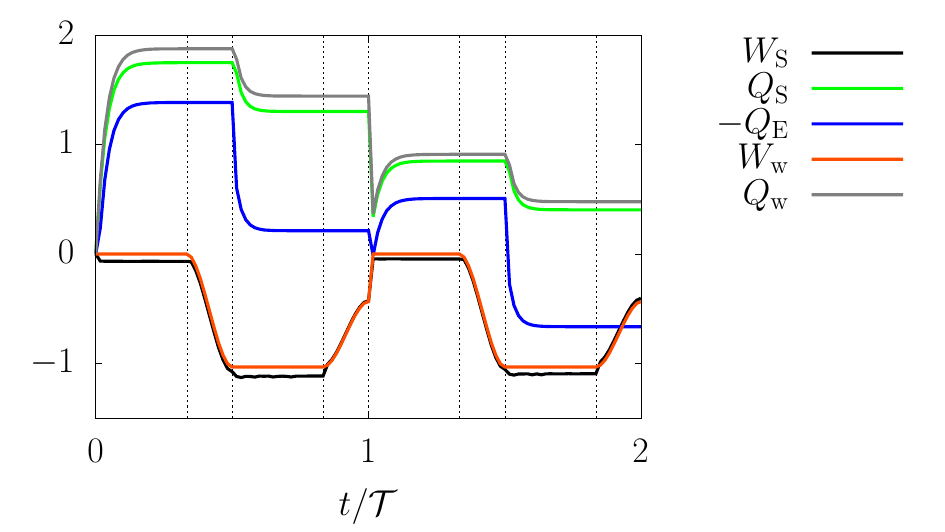} 
	\caption{Work $W_{\rm S}(t)$ and heat $Q_{\rm S}(t)$ as a function of time, as well as a comparison with the weak coupling work $W_{\rm w}(t)$, heat $Q_{\rm w}(t)$ and the heat contribution from the environment $-Q_{\rm E}(t)$. Dotted vertical lines indicate the boundaries between strokes of the Otto cycle. Parameters, given in units of $\Gamma$, are $U=2$, $\varepsilon_{\rm h} =2$, $\varepsilon_{\rm c} =0.6$, $k_{\rm B}T_{\rm h}=10$, $k_{\rm B}T_{\rm c}=1$. }
	\label{termoabove0}
\end{figure}


In this section, we analyze the results for a case where the energy levels are above the Fermi level ($\varepsilon_{\rm h} ,\varepsilon_{\rm c} > 0$), considering both the noninteracting ($U=0$) and interacting ($U\neq 0$) parameter regime. The corresponding energy-level scheme is shown in Fig.~\ref{fig:energy_levels} (left panel).

Fig.~\ref{popabove1} shows the population of the different states of the reduced density matrix $\rho_{\rm S}(t)$ as a function of time. These results, obtained for parameters $\varepsilon_{\rm h} = 2$, $\varepsilon_{\rm c} = 0.6$, and $k_{\rm B} T_{\rm h} = 10$, $k_{\rm B} T_{\rm c} = 1$, indicate that the system evolves fast towards a steady-state regime. For both the noninteracting ($U=0$, dashed lines) and interacting ($U=2$, solid lines) cases, the system reaches a steady-state limit cycle after just two iterations. In the noninteracting case, the system effectively behaves as two decoupled single-dot systems, exhibiting characteristic population dynamics: the singly and doubly occupied states are populated during the hot stroke and subsequently depopulated during the cold stroke. In contrast, in the interacting case, the Coulomb interaction pushes the doubly occupied energy level further away from the Fermi level, making these transitions less probable; as a result, the doubly occupied state remains almost depopulated after Stroke~4, while the dynamics of the singly occupied states are only weakly affected.

Fig.~\ref{termoabove1} depicts work $W_{\rm S}(t)$ and heat $Q_{\rm S}(t)$ as functions of time for the same setup, comparing the noninteracting ($U=0$, dashed lines) and interacting ($U=2$, solid lines) cases. The work $W_{\rm S}(t)$ reflects both the energy required to renormalize the system’s energy levels due to system-environment coupling during strokes I and III, as well as the energy associated with externally shifting the levels in strokes II and IV. The heat $Q_{\rm S}(t)$, on the other hand, accounts for the energy exchanged with the environment. It is positive during stroke I, corresponding to energy absorption used to populate the singly and doubly occupied states, and negative during stroke III, as energy is released while these states are depopulated. While the overall work output is similar in both cases, the Coulomb interaction increases the heat $Q_{\rm S}$, indicating that more heat is required to achieve the same work output as in the noninteracting case.

Fig.~\ref{effabove1} compares the three definitions of efficiency (Eqs.~\eqref{def-etah}--\eqref{def-etaelb}) 
for both the interacting case ($U=2$, solid lines) and the noninteracting case ($U=0$, dashed lines). 
In all definitions, the presence of Coulomb interaction reduces the cycle efficiency, 
as more heat is required to achieve a similar work output. 
The relation $\eta_{\rm h}^w \leq \eta_{\rm h} \leq \eta_{\rm h}^{\mathrm{ELB}}$ is generally observed in our simulations, which can be interpreted as reflecting, at least partially, the contribution of the interaction energy $\langle H_{\mathrm{int}} \rangle$ in $\langle K_{\rm S} \rangle$, whereas $\eta_{\rm h}^w$ does not include this contribution. A more detailed discussion of the role of $\langle H_{\mathrm{int}} \rangle$ in the minimal dissipation approach is provided in Appendix~\ref{sec:compare}.

Finally, Fig.~\ref{termoabove0} shows the time evolution of work $W_{\rm S}(t)$ and heat $Q_{\rm S}(t)$ for the interacting case ($U=2$), together with the weak-coupling quantities $W_{\rm w}(t)$, $Q_{\rm w}(t)$, and the environmental heat contribution $-Q_{\rm E}(t)$. The results indicate that $W_{\rm S}$ deviates slightly from $W_{\rm w}$ due to the influence of the environment, which performs some work on the system. Correspondingly, $Q_{\rm S}$ lies between $Q_{\rm w}$ and $-Q_{\rm E}$. Here, $Q_{\rm w}$ is the weak-coupling heat, which accounts for the environmental action but neglects $\langle H_{\mathrm{int}} \rangle$, whereas $Q_{\rm S}$ includes it partially through $\langle K_{\rm S} \rangle$, and $-Q_{\rm E}$ represents the heat exchanged purely with the environments.

\subsection{Energy levels below the Fermi level}\label{sec:below}

We now analyze the case where both energy levels lie below the Fermi energy ($\varepsilon_{\rm h},\varepsilon_{\rm c}<0$), considering both the noninteracting ($U=0$) and interacting ($U \neq 0$) cases. The corresponding energy-level scheme is shown in Fig.~\ref{fig:energy_levels} (right panel).

\begin{figure}[t]
	\centering
	\includegraphics[width=0.50\textwidth]{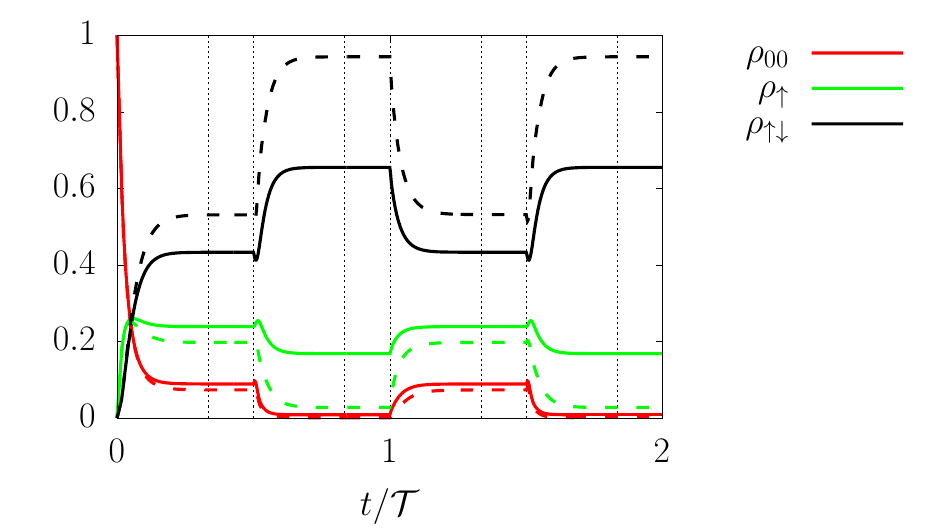} 
	\caption{Population of the different states of $\rho_{\rm S}(t)$ as a function of time for the noninteracting case ($U=0$, dashed lines) and the interacting case ($U=2$, solid lines). Dotted vertical lines indicate the boundaries between strokes of the Otto cycle. Parameters, given in units of $\Gamma$, are $\varepsilon_{\rm h}=-5$, $\varepsilon_{\rm c}=-2.8$, $k_{\rm B}T_{\rm h}=10$, $k_{\rm B}T_{\rm c}=1$.}
	\label{popbelow1}
\end{figure}

Fig.~\ref{popbelow1} shows the population of the different states of the reduced density matrix $\rho_{\rm S}(t)$ as a function of time. 
For the noninteracting case ($U=0$, dashed lines), during the hot stroke, transitions from the unoccupied state to the singly and doubly occupied states are energetically favorable. In contrast, transitions from the singly or doubly occupied states back to lower-occupation states are energetically suppressed. This asymmetry in transition probabilities results in a steady-state distribution dominated by the doubly occupied state, followed by the singly occupied states. During the cold stroke, this asymmetry toward the doubly occupied state becomes even more pronounced. This is due to the sharper profile of the Fermi distribution at lower temperatures, which further enhances the probability of energetically favorable transitions into the doubly occupied state, while suppressing the corresponding reverse processes.    

When Coulomb interaction is present ($U=2$, solid lines), it effectively lowers the energy difference between the doubly occupied state and the Fermi level. Transitions to the singly and unoccupied states become more likely, reducing the population of the doubly occupied state compared to the noninteracting case. Although the cold stroke still drives the system toward the doubly occupied state, its dynamics are now affected by this interaction-induced modification, which increases transition rates compared to the case where the energy levels lie above the Fermi energy.

\begin{figure}[t]
	\centering
	\includegraphics[width=0.50\textwidth]{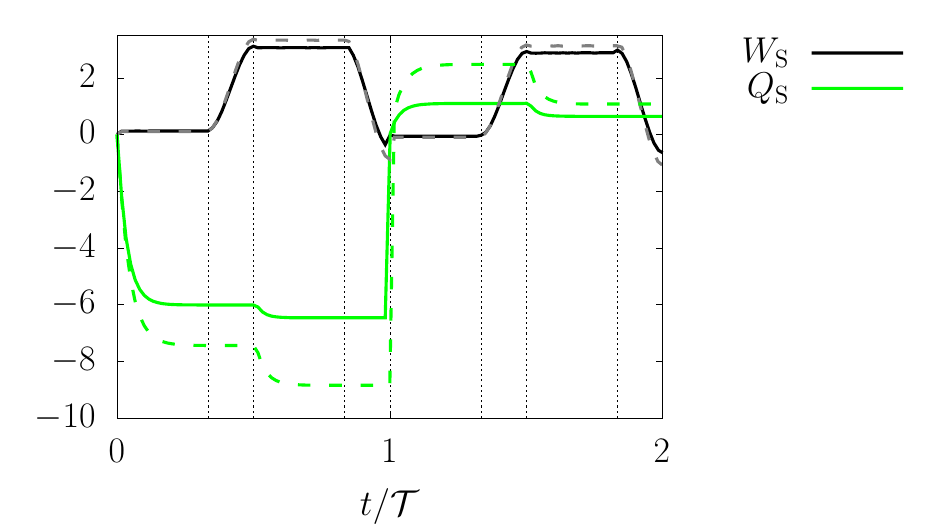} 
	\caption{Work $W_{\rm S}(t)$ and heat $Q_{\rm S}(t)$ as a function of time, comparing the noninteracting case ($U=0$, dashed lines) with the interacting case ($U=2$, solid lines). Dotted vertical lines indicate the boundaries between strokes of the Otto cycle. Parameters, given in units of $\Gamma$, are $\varepsilon_{\rm h}=-5$, $\varepsilon_{\rm c}=-2.8$, $k_{\rm B}T_{\rm h}=10$, $k_{\rm B}T_{\rm c}=1$.}
	\label{termobelow1}
\end{figure}

Fig.~\ref{termobelow1} shows the time evolution of work $W_{\rm S}(t)$ and heat $Q_{\rm S}(t)$ for the same set of parameters. 
In the noninteracting case ($U=0$, dashed lines), each cycle iteration produces a negative net work output, while $Q_{\rm S}$ remains positive once the limit cycle is reached.
When Coulomb interaction is included ($U=2$, solid lines), the cycle still operates in the same thermodynamic regime, but the net work output $W_{\rm S}$ is reduced compared to the noninteracting case. 
At the same time, the heat released during the cold stroke is also reduced, as evidenced by the smaller decrease in $Q_{\rm S}$.

\begin{figure}[t]
	\centering
	\includegraphics[width=0.50\textwidth]{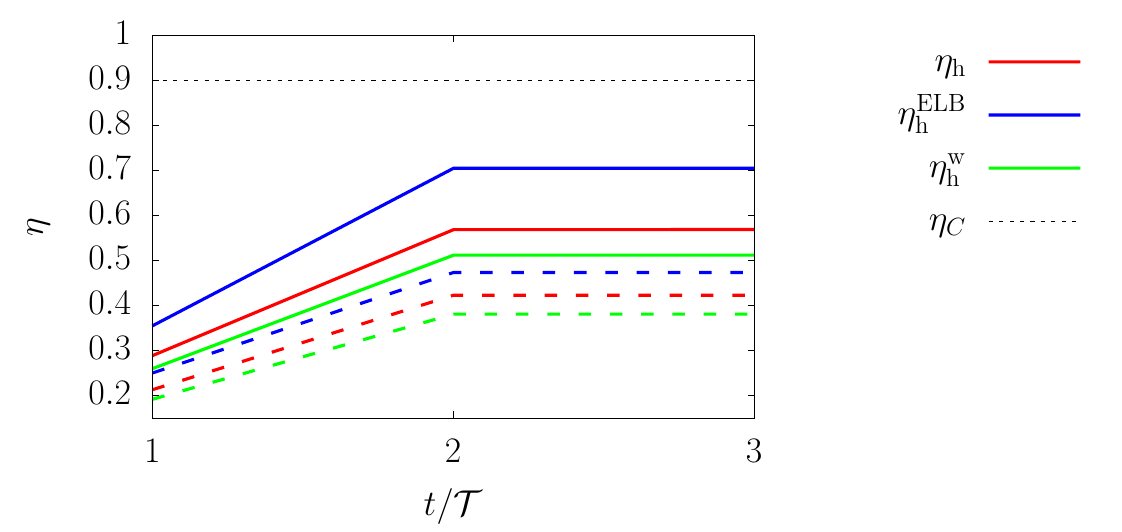} 
	\caption{Comparison of the three efficiency definitions (Eqs.~\eqref{def-etah}--\eqref{def-etaelb}) 
		for the interacting case ($U=2$, solid lines) and the noninteracting case ($U=0$, dashed lines). 
		The black dotted line indicates the Carnot efficiency $\eta_C$.  
		Parameters, given in units of $\Gamma$, are $\varepsilon_{\rm h}=-5$, $\varepsilon_{\rm c}=-2.8$, $k_{\rm B}T_{\rm h}=10$, $k_{\rm B}T_{\rm c}=1$. }
	\label{effbelow1}
\end{figure}

Fig.~\ref{effbelow1} shows a comparison of the three efficiency definitions (Eqs.~\eqref{def-etah}--\eqref{def-etaelb}) for the interacting case ($U=2$, solid lines) and the noninteracting case ($U=0$, dashed lines).  
All definitions predict that the interacting cycle can achieve a higher efficiency than the noninteracting cycle, which is consistent with the reduced heat required to produce similar work due to the Coulomb interaction effectively bringing the doubly occupied state closer to the Fermi level.  
As in the previous comparison, $\eta_{\rm h}$ lies between $\eta_{\rm h}^w$ and $\eta_{\rm h}^{\mathrm{ELB}}$, due to the partial inclusion of the interaction energy in $\langle K_{\rm S} \rangle$, whereas the weak-coupling definition $\eta_{\rm h}^w$ does not include this contribution.

\begin{figure}[t]
	\centering
	\includegraphics[width=0.50\textwidth]{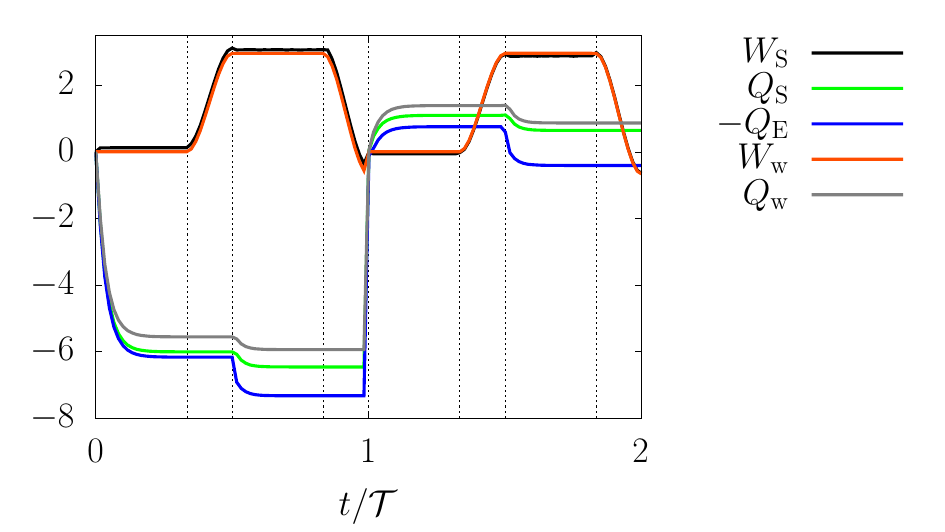} 
	\caption{Work $W_{\rm S}(t)$ and heat $Q_{\rm S}(t)$ compared with the weak-coupling work $W_{\rm w}(t)$, heat $Q_{\rm w}(t)$ and the environmental heat contribution $-Q_{\rm E}(t)$. Dotted vertical lines indicate the boundaries between strokes of the Otto cycle. Parameters: $U=2$, $\varepsilon_{\rm h}=-5$, $\varepsilon_{\rm c}=-2.8$, $k_{\rm B}T_{\rm h}=10$, $k_{\rm B}T_{\rm c}=1$. Parameters and results are given in units of $\Gamma$.}
	\label{termobelow0}
\end{figure}

Finally, Fig.~\ref{termobelow0} presents $W_{\rm S}(t)$ and $Q_{\rm S}(t)$ as a function of time, together with the weak-coupling expressions $W_{\rm w}(t)$, $Q_{\rm w}(t)$, and the environmental heat contribution $-Q_{\rm E}(t)$.  
As in the previous case, $W_{\rm S}$ deviates slightly from $W_{\rm w}$ due to the influence of the environment, which performs some work on the system. Correspondingly, $Q_{\rm S}$ lies between $Q_{\rm w}$ and $-Q_{\rm E}$, reflecting the partial incorporation of the interaction energy contribution in $\langle K_{\rm S} \rangle$, while $Q_{\rm w}$ neglects it.

\section{Conclusions}\label{sec:conclu}

In this work, we have presented a comprehensive study of the thermodynamics of a periodic quantum Otto cycle implemented on a single-impurity Anderson model. The quantum dynamics of the system was described employing the hierarchical equations of motion (HEOM) approach, which provides a numerically exact treatment that is particularly crucial for exploring the impact of strong system-reservoir coupling and Coulomb interactions. To analyze the thermodynamic properties of the cycle process, the study uses definitions of work and heat based on the principle of minimal dissipation. The findings are compared to other definitions of work and heat considered recently.

Our analysis revealed that Coulomb interactions introduce significant modifications to the thermodynamic performance of the cycle, notably shifting operational regimes in the energy parameter space and affecting both the heat exchanges and work output. In particular, the presence of interaction leads to a nontrivial enhancement of efficiency in specific regimes where population dynamics of doubly occupied states favorably influence the cycle’s energy balance. This finding, which is obtained with all thermodynamic definitions employed, highlights the potential of exploiting strong correlations to improve quantum thermal machines beyond classical limits.

Future work may extend this framework to multi-level impurities or more complex reservoir engineering to explore novel operational modes and optimization strategies for nanoscale quantum engines. Moreover, finite time performance and its impact on the cycle here described could also be addressed with these methods.

\acknowledgments

This project was supported by the European Union's Framework Programme 
for Research and Innovation Horizon 2020 (2014-2020) under the 
Marie Sk\l{}odowska-Curie Grant Agreement No.~847471 , and by the DFG fund via the Research Training Group "Dynamics of Controlled Atomic and Molecular Systems" (RTG 2717), and the Research Unit "Reducing Complexity of Nonequilibrium Systems" (FOR 5099).  Furthermore, support by the state of Baden-Württemberg through
bwHPC and the DFG through Grant No. INST 40/575-1
FUGG (JUSTUS 2 cluster) is gratefully acknowledged. A.C. acknowledges further support from MUR via the PRIN 2022 Project “Quantum Reservoir Computing (QuReCo)” (Contract No. 2022FEXLYB). M.T. thanks David Gelbwaser-Klimovsky and Uri Peskin for many interesting discussions and their great hospitality at the Technion in Haifa. This work was supported in part at the Technion by a fellowship from the Lady Davis foundation.
\appendix

\section{Interaction energy contributions to $\langle K_{\rm S} (t) \rangle$}
\label{sec:compare}

In this section, we explore the relationship between the expectation value of the effective Hamiltonian, denoted as $\langle K_{\rm S}(t) \rangle$, and the expectation values of the bare Hamiltonian $\braket{H_{\rm S}}$ and the interaction Hamiltonian $\braket{H_\mathrm{int}}$, providing insight into how much the interaction energy is accounted for in the internal energy of the system. 
To do this, we perform a perturbative expansion of $\langle K_{\rm S}(t) \rangle$ in terms of the coupling strength $\lambda= \Gamma^{1/2}$.

Let us assume for ease of notation that the interaction Hamiltonian is written as $H_\mathrm{int} = \lambda A \otimes B$, where $A$ and $B$ are operators on the system and environment side, respectively. The extension to the general case of $H_\mathrm{int} = \lambda \sum_i A_i \otimes B_i$ is straightforward and does not affect the main message of this section. We then move to the interaction picture, where the Hamiltonian is written as $\tilde{H}_t = \lambda \tilde A_t \otimes \tilde B_t$, with $\tilde{H}_t = e^{i(H_{\rm S}+H_{\rm E})t}H_\mathrm{int}e^{-i(H_{\rm S}+H_{\rm E})t}$, $\tilde{A}_t = e^{iH_{\rm S} t}A e^{-iH_{\rm S} t}$, and $\tilde{B}_t = e^{iH_{\rm E} t}B e^{-iH_{\rm E} t}$.

The evolution of the entire system in interaction picture, $\tilde \rho_t = e^{i(H_{\rm S}+H_{\rm E})t}\rho_{\rm S+E}(t) e^{-i(H_{\rm S}+H_{\rm E})t}$ (note that we dropped the subscript $\rm S+E$ on the LHS for ease of notation), can be written in orders of $\lambda$ by iterating the integral representation of the von Neumann evolution,
\begin{equation}
\tilde{\rho}_t = \tilde{\rho}_0 -i \lambda \int_0^t d\tau [\tilde A_\tau \otimes \tilde B_\tau, \tilde{\rho}_\tau] \; ,
\end{equation}
which gives, up to first order in $\lambda$ and for factorizing initial conditions $\rho_{\rm S+E}(0)= \rho_{\rm S}\otimes \rho_{\rm E}$,
\begin{equation}
\tilde{\rho}^{(1)}_t = \rho_{\rm S}\otimes\rho_{\rm E} -i \lambda \int_0^t d\tau [\tilde A_\tau \otimes \tilde B_\tau, \rho_{\rm S}\otimes\rho_{\rm E}] \; .
\end{equation}
We can now make use of the perturbation expansion of $K_{\rm S}(t)$ provided in \cite{Colla2025Ks,Colla2025TCL}. Its first order contribution in interaction picture reads
\begin{align}
	\tilde K_{\rm S,t}^{[1]}=\braket{\tilde B_t}_0\left[ \tilde A_t - \braket{\tilde A_t}_{\mathrm{m}}\right] \; .
\end{align}
Here, we introduced the following notation:
\begin{align}
	\braket{\tilde B_t}_0 &= \Tr_{\rm E} \{ \tilde B_t \, \rho_{\rm E} \}, \\
	\braket{\tilde A_t}_0 &= \Tr_{\rm S} \{ \tilde A_t \, \rho_{\rm S} \}, \\
	\braket{\tilde A_t}_{\rm m} &= \frac{1}{d} \Tr_{\rm S} \{ \tilde A_t \, \mathbb{I} \} \; ,
\end{align}
where $d$ is the dimension of the system Hilbert space ($d=4$ in our case).  
In particular, $\braket{\tilde A_t}_{\rm m}$ corresponds to the expectation value of $\tilde A_t$ if the system is initially in the maximally mixed state.

This last contribution to $\tilde K_{\rm S,t}$ comes from the requirement that the effective Hamiltonian be traceless.
Then, to obtain an approximation of $\braket{K_{\rm S}(t)}$ valid up to first order, we add the expansion of the first order contribution (for simplicity, still calculated in interaction picture) to $\braket{H_{\rm S}}$, which contains the zeroth order contribution and the higher order ones due to the evolution of $\rho_{\rm S}(t)$:
\begin{equation}
\braket{K_{\rm S}(t)} = \braket{H_{\rm S}} + \lambda \braket{\tilde B_t}_0 \left[ \braket{\tilde A_t}_0 -  \braket{\tilde A_t}_{\mathrm{m}}\right] + \mathcal{O}(\lambda^2)  \; ,
\end{equation}
while the interaction energy is given by
\begin{equation}
\braket{H_\mathrm{int}} = \lambda  \braket{\tilde B_t}_0 \braket{\tilde A_t}_0 + \mathcal{O}(\lambda^2)  \;.
\end{equation}
Then, introducing $\langle H_\mathrm{int} \rangle_{\mathrm{m}}$ as the expectation value of the interaction Hamiltonian $H_\mathrm{int}$ at time $t$ when the system starts in the maximally mixed state, and defining
\begin{align} \delta \langle H_\mathrm{int} \rangle := \langle H_\mathrm{int} \rangle - \langle H_\mathrm{int} \rangle_{\mathrm{m}} , \end{align}
we find
\begin{align} \langle K_{\rm S} (t) \rangle = \braket{H_{\rm S}} + \delta\braket{H_\mathrm{int}} + O(\Gamma) ,  \end{align}
namely at first order the effective internal energy takes into account part of the system-environment interaction energy (an amount only due to the fact that the initial system state is not maximally mixed).

If the first order contribution vanishes (which happens whenever $\braket{\tilde B_t}_0=0$, and which is the case for the model studied in this work) we consider the second order contribution. In this case, the interaction energy is given by
\begin{align}
\braket{H_\mathrm{int}} =  -i \lambda^2 \int_0^t d\tau \big[ &\braket{\tilde A_t \tilde A_{\tau}}_{0} \braket{\tilde B_t \tilde B_{\tau}}_0 -   \text{h.c.} \big] + \mathcal{O}(\lambda^3)  \;,
\end{align}
where we defined
\begin{align}\label{eq:app_AA}
    \braket{\tilde A_t \tilde A_{\tau}}_{0} &= \Tr\{\tilde A_t \tilde A_\tau \rho_{\rm S}\} \; , \\
    \braket{\tilde B_t \tilde B_{\tau}}_{0} &= \Tr\{\tilde B_t \tilde B_\tau \rho_{\rm E}\} \; .
\end{align}

The first order contribution of the effective Hamiltonian vanishes, while the second order reads, assuming also $\Tr\{\tilde A_t\}=0$ \cite{Colla2025TCL}:
\begin{align}
    \tilde K_{\rm S,t}^{[2]} (t) = \frac{1}{2i} \int_0^t d\tau \left(\braket{\tilde B_t \tilde B_{\tau}}_0 \left[ \tilde A_t \tilde A_{\tau} - \braket{\tilde A_t \tilde A_{\tau}}_{\mathrm{m}} \right] - \text{h.c.}\right) \; ,
\end{align}
where, as before, the notation $\braket{\tilde A_t \tilde A_{\tau}}_{\mathrm{m}}$ indicates expression \eqref{eq:app_AA} taken with respect to the maximally mixed state of the system.

By taking the expectation value with respect to the zeroth order in the system state, and adding the contribution of $\braket{H_S}$, the above leads to
\begin{align} \langle K_{\rm S} (t) \rangle = \braket{H_{\rm S}} + \frac{1}{2}\delta\braket{H_\mathrm{int}} + O(\Gamma^{3/2}) .  \end{align}
In this expression, the factor of ${1}/{2}$ appears due to the specific structure of the perturbative expansion when the first-order term vanishes.

\begin{figure}[t]
	\includegraphics[scale=0.50]{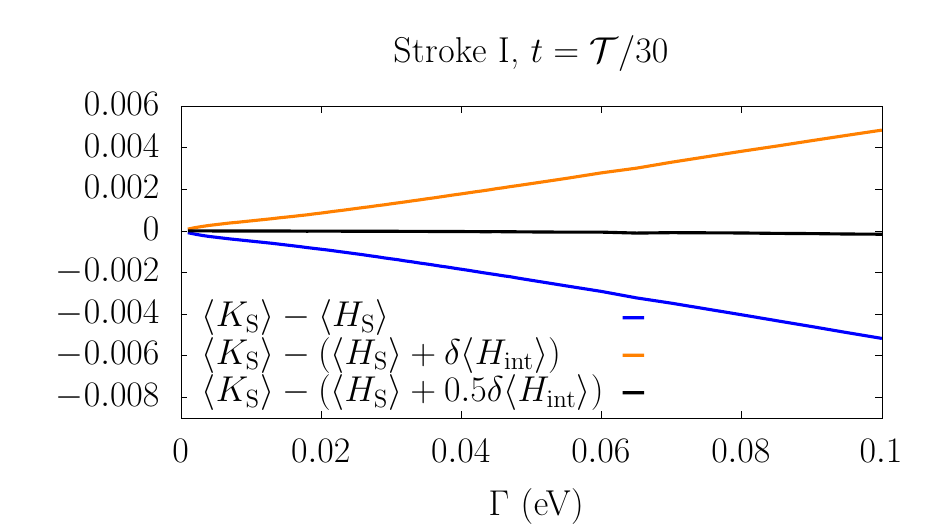}
	\caption{Difference between the expectation value of the effective Hamiltonian, $\langle K_{\rm S} \rangle$, and different orders of the perturbative expansion  as defined in the text. Results are shown in terms of the coupling strength $\Gamma$, for the non-interacting case treated in Sec. \ref{sec:above}, at the beginning of Stroke I.}
	\label{comparebegin}
\end{figure}
\begin{figure}[t]
	\includegraphics[scale=0.50]{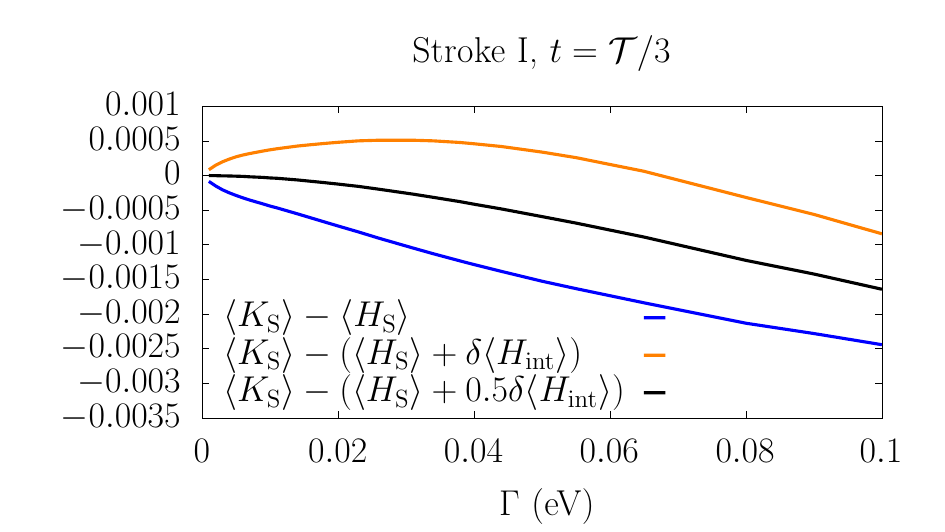}
	\caption{Difference between the expectation value of the effective Hamiltonian, $\langle K_{\rm S} \rangle$, and different order of the perturbative expansion as defined in the text. Results are shown in terms of the coupling strength $\Gamma$, for the non-interacting case treated in Sec. \ref{sec:above}, at the end of Stroke I.}
	\label{compareend}
\end{figure}

Now, we proceed to discuss the specific comparison between $\langle K_{\rm S} (t)\rangle$ and the different orders of perturbative expansion for the noninteracting case treated in Sec. \ref{sec:above}, where the Coulomb interaction is set to zero, and the system's energy levels are situated above the Fermi level. This analysis is done in the context of the hot phase of the limit cycle.

In Fig.~\ref{comparebegin}, we show the difference between the expectation value of the effective Hamiltonian $\langle K_{\rm S} (t)\rangle$ and the various orders of the perturbative expansion, at the beginning of Stroke I. The key observation here is that at the very beginning of the hot phase, $\langle K_{\rm S} (t) \rangle$ is well approximated by $\braket{H_{\rm S}} + 0.5 \delta \braket{H_\mathrm{int}}$, meaning that the second-order perturbative correction contributes significantly to the effective Hamiltonian's expectation value.

As time progresses during the hot phase, we observe a deviation between $\langle K_{\rm S} (t) \rangle$ and $\braket{H_{\rm S}} + 0.5 \delta \braket{H_\mathrm{int}}$. This is illustrated in Fig.~\ref{compareend}, where the same differences are shown at the end of Stroke I. Here, we can see that as the system evolves, the higher-order terms in the perturbative expansion become increasingly important, and the coupling strength $\Gamma$ determines the magnitude of the deviation.

\section{Analytical results in the weak coupling case}
\label{sec:weekcoupling}

\begin{figure*}[htbp]
	\centering
	\includegraphics[width=0.3\textwidth]{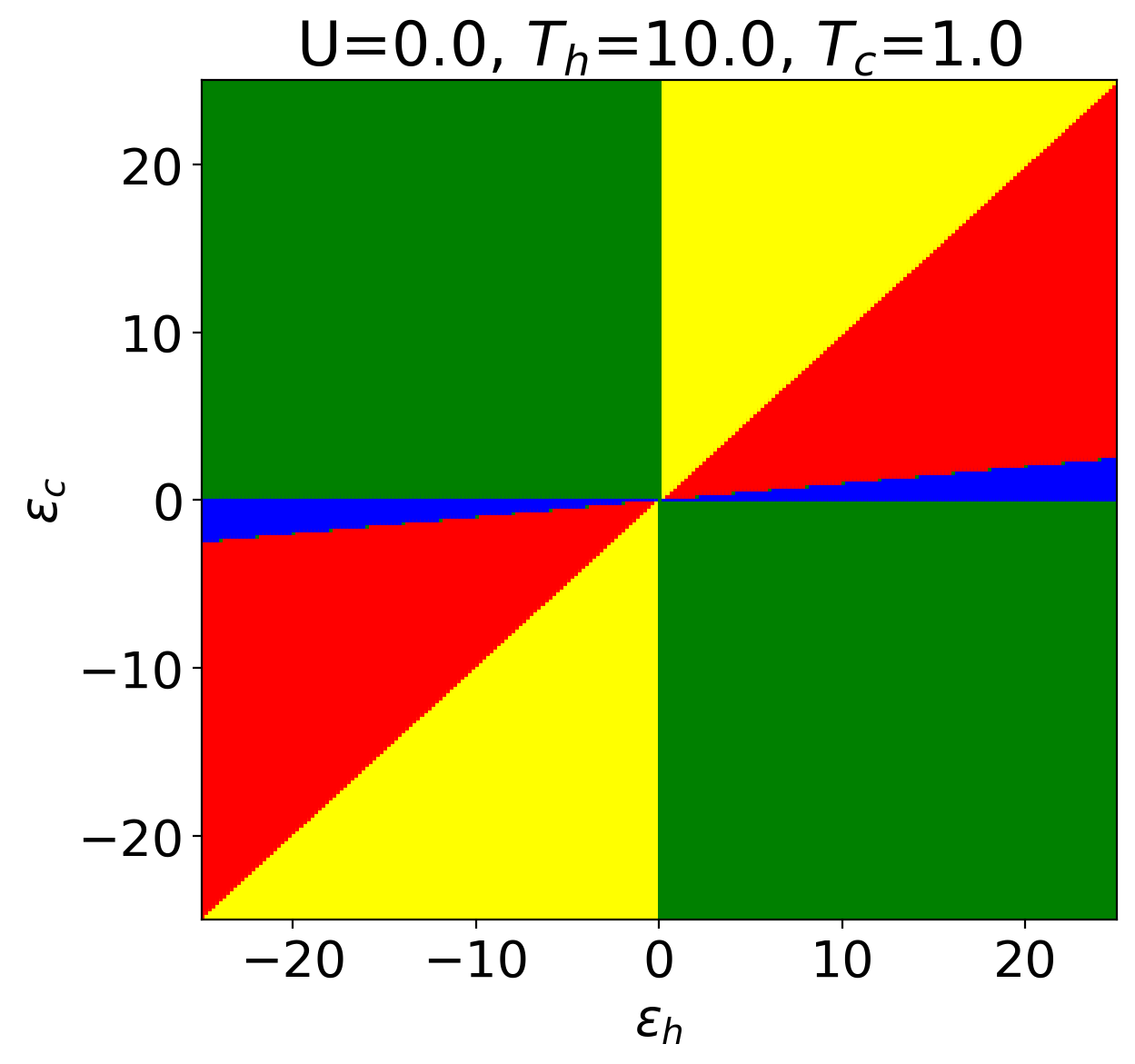}\hfill
	\includegraphics[width=0.3\textwidth]{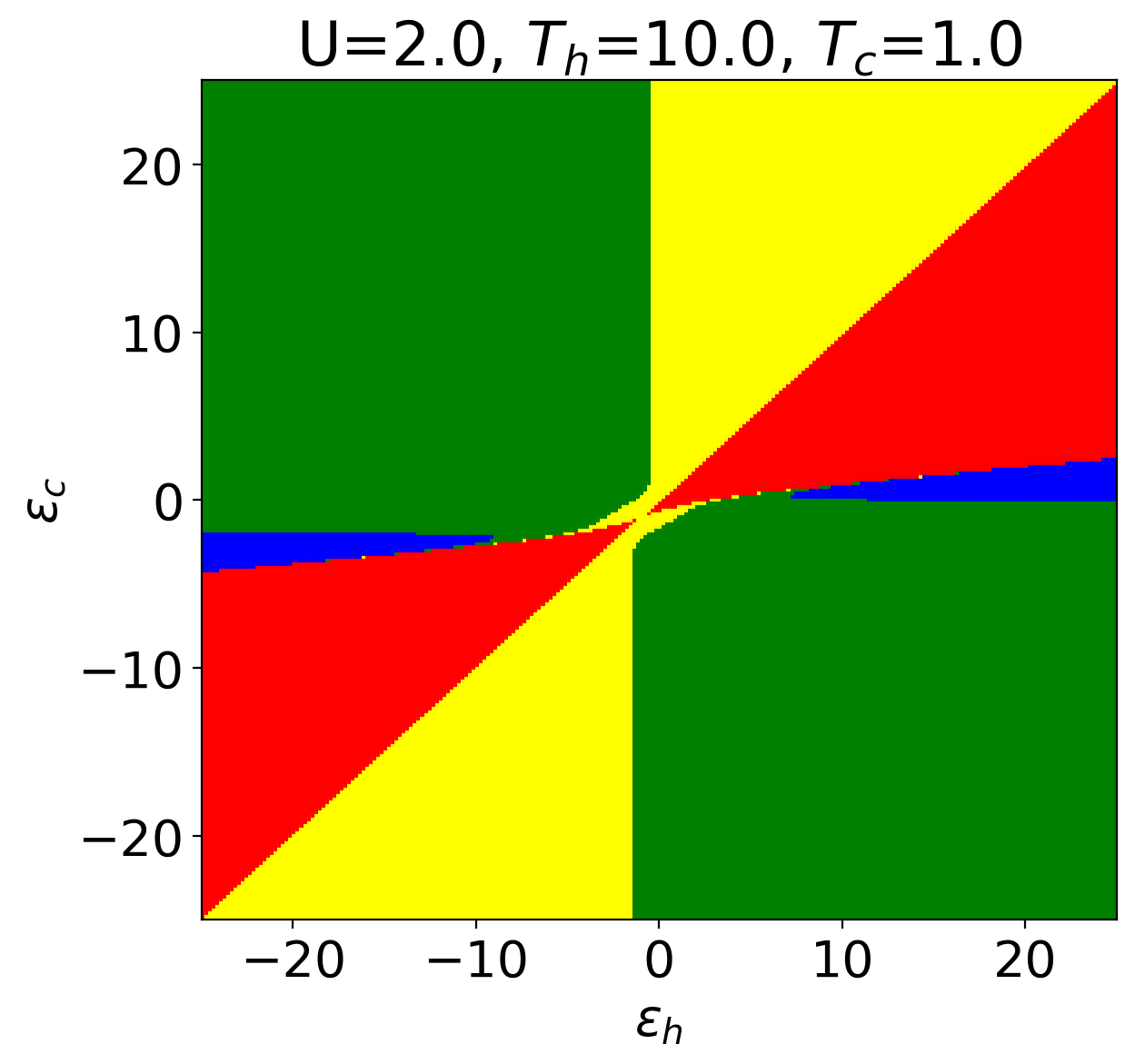}\hfill
	\includegraphics[width=0.3\textwidth]{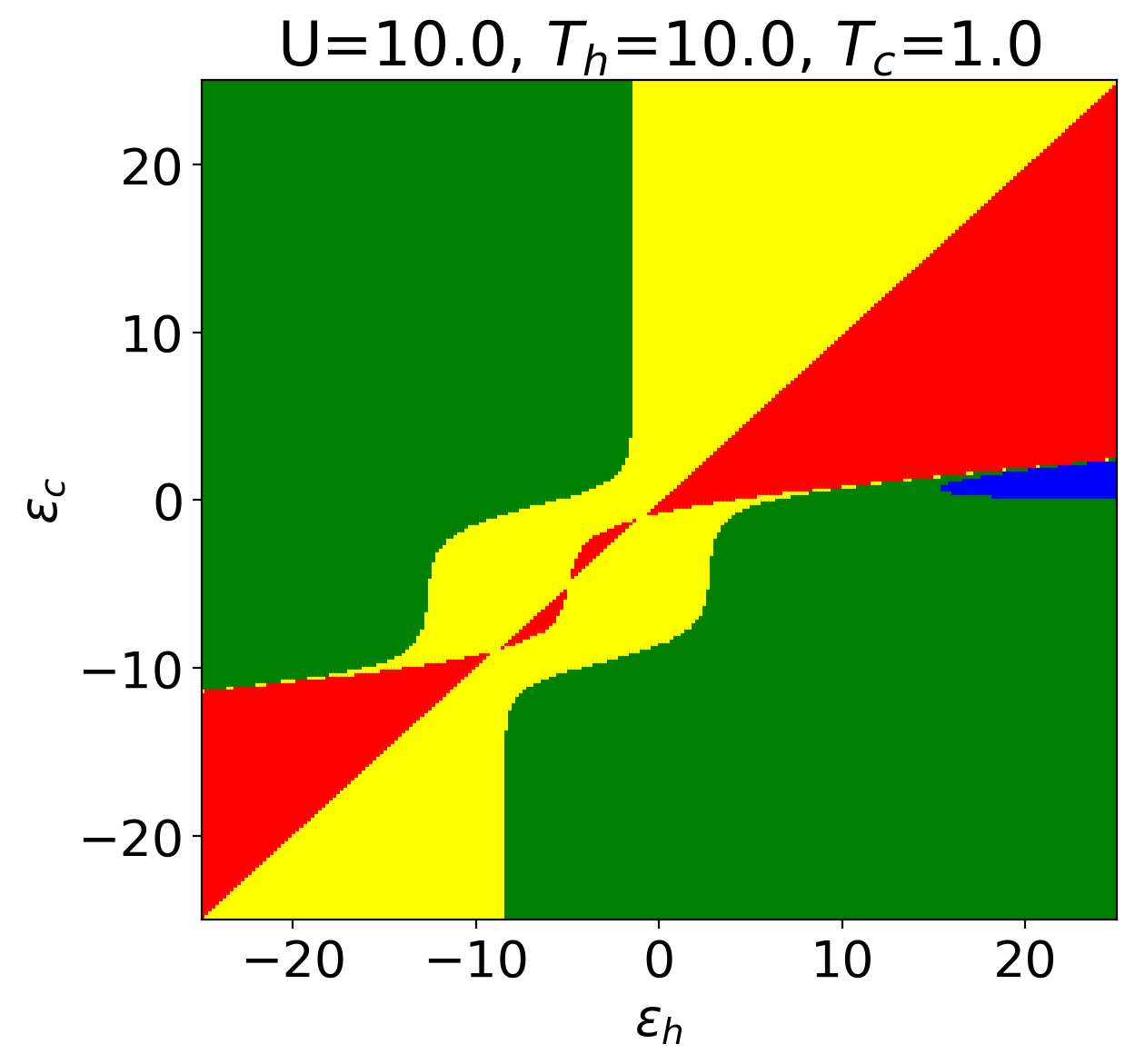}
	
	\vspace{0.5cm}
	
	\includegraphics[width=0.3\textwidth]{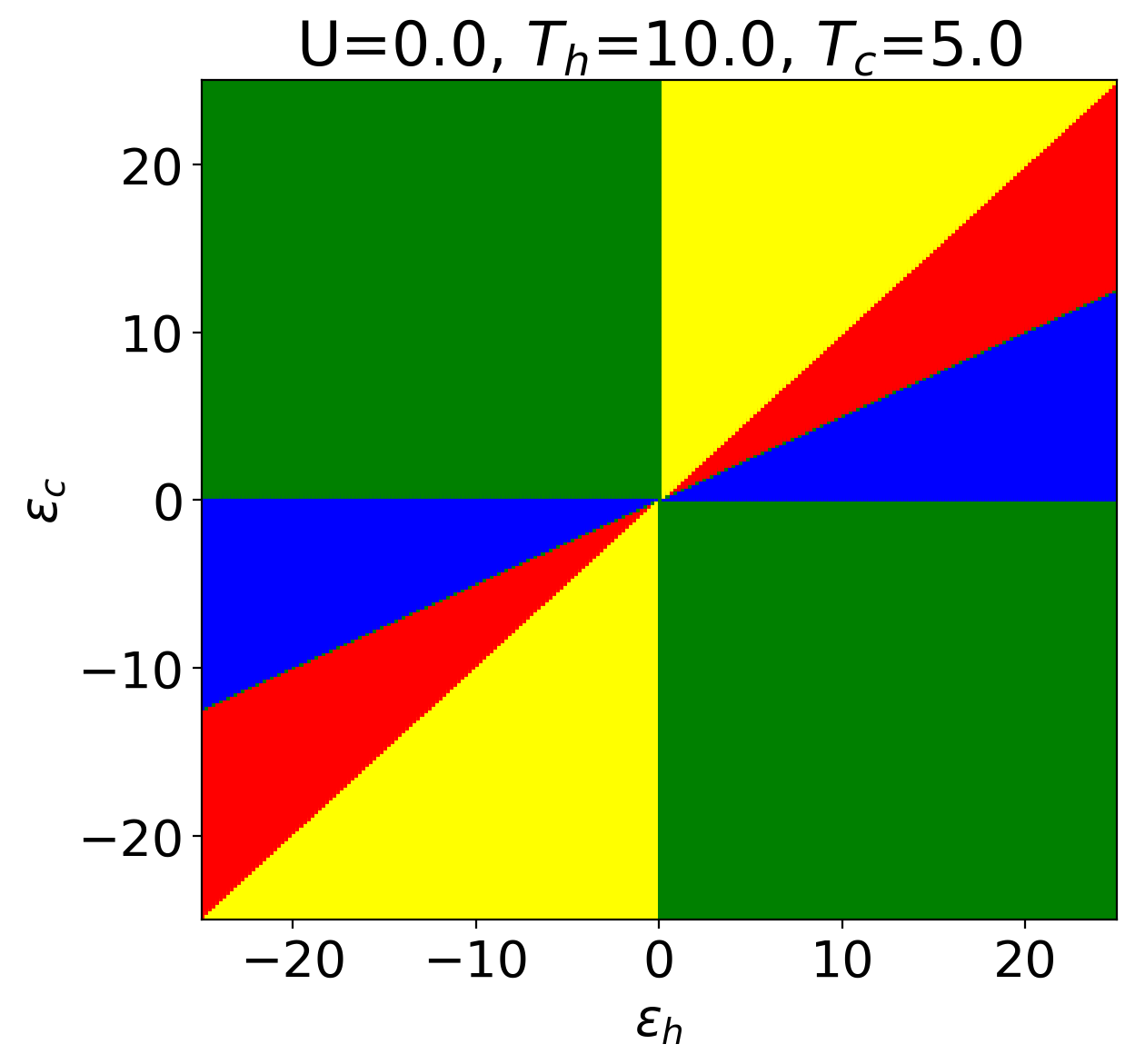}\hfill
	\includegraphics[width=0.3\textwidth]{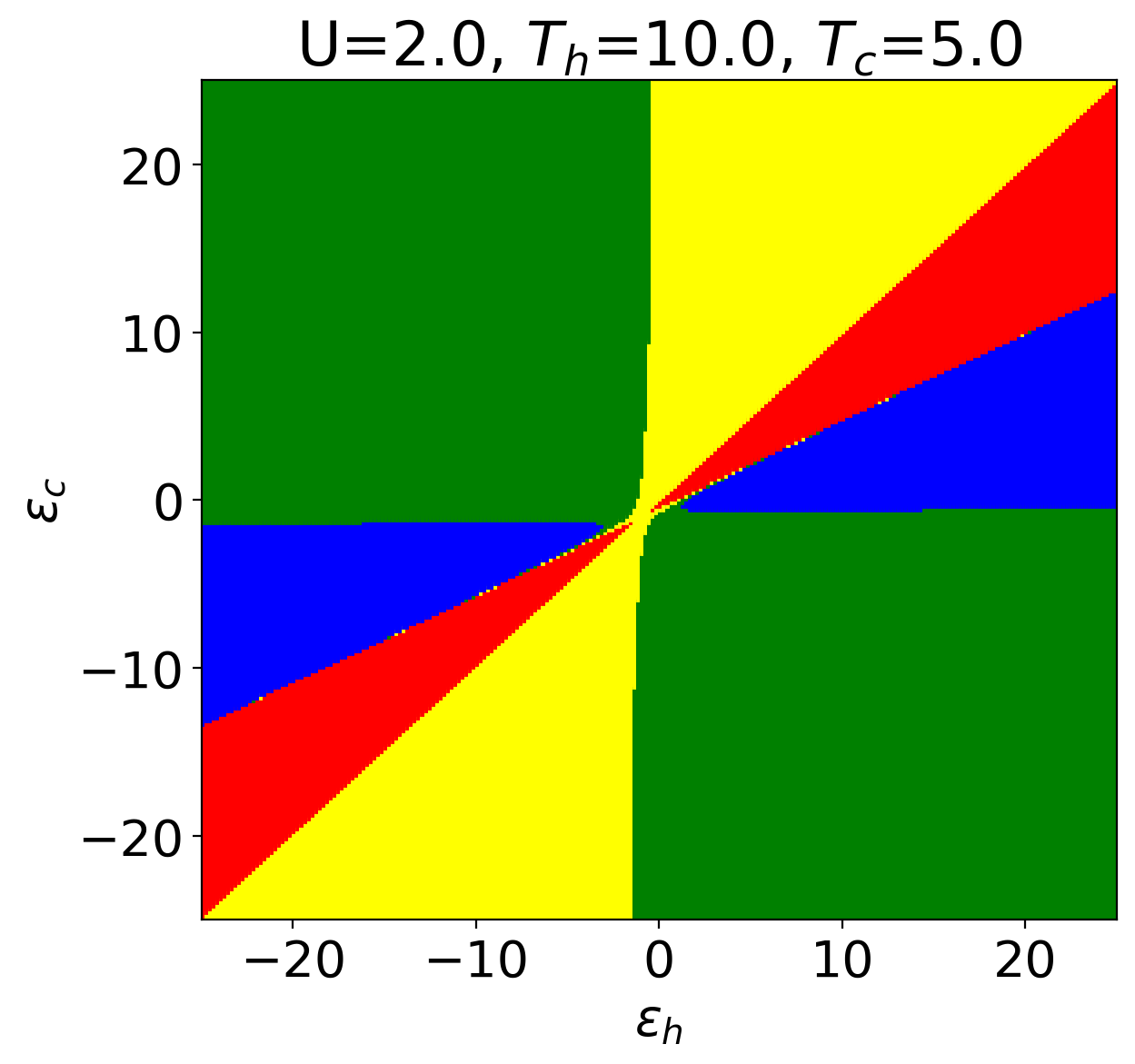}\hfill
	\includegraphics[width=0.3\textwidth]{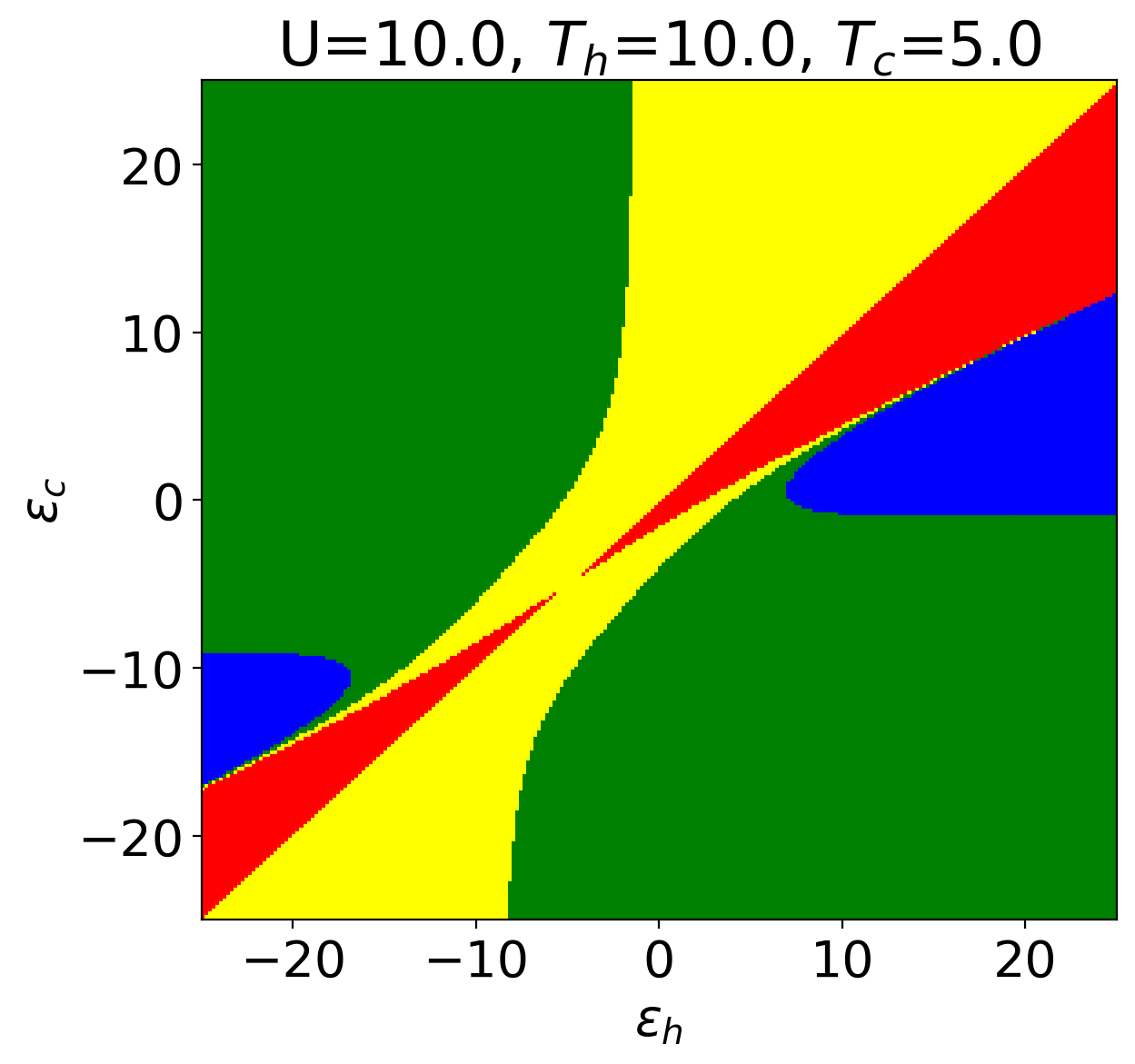}
	
	\caption{Operating regimes of the quantum Otto cycle as a function of $\varepsilon_{\rm h}$ (horizontal axis), $\varepsilon_{\rm c}$ (vertical axis) in the regime of weak system-environment coupling $\Gamma$. Different colors indicate different operating regimes: heat engine (red), refrigerator (blue), accelerator (yellow), and heater (green). The figures illustrate how these regimes vary with temperature and Coulomb interaction $U$.  Parameters are given in units of $\Gamma$.}
	\label{fig:otto_working_regimes}
\end{figure*}

In this section, we analytically investigate the thermodynamic performance of the quantum Otto cycle under the weak coupling approximation for the system-environment interaction. Starting from the system Hamiltonian in Eq.~\eqref{eq:general_Hamiltonian}, we consider a perturbative treatment of the system--reservoir interaction, valid in the regime $\Gamma \ll k_{\rm B} T_{\rm h,c}$, where the system dynamics can be accurately captured by a Markovian quantum master equation in the form of rate equations. Following the approach developed by Gurvitz and Prager~\cite{Gurvitz1996}, the population dynamics of the reduced density matrix, when coupled to a single reservoir labeled by $\rm K = h, c$, is governed by the set of equations:
\begin{align*}
	\frac{\partial}{\partial t} \rho_{00} &= -2\Gamma f_{\rm K}\, \rho_0 + 2\Gamma (1 - f_{\rm K})\, \rho_\uparrow, \\
	\frac{\partial}{\partial t} \rho_\uparrow &= \Gamma f_{\rm K}\, \rho_0 - \Gamma (1 - f_{\rm K} + f_{\rm K}^U)\, \rho_\uparrow + \Gamma (1 - f_{\rm K}^U)\, \rho_{\uparrow\downarrow}, \\
	\frac{\partial}{\partial t} \rho_{\uparrow\downarrow} &= 2\Gamma f_{\rm K}^U\, \rho_\uparrow - 2\Gamma (1 - f_{\rm K}^U)\, \rho_{\uparrow\downarrow},
\end{align*}
where $\rho_0$, $\rho_\uparrow$, and $\rho_{\uparrow\downarrow}$ denote the populations of the empty, singly occupied, and doubly occupied states, respectively. The Fermi-Dirac distribution functions entering the transition rates are defined as
\begin{align*}
	f_{\rm K} &= \left( e^{\beta_{\rm K} \varepsilon} + 1 \right)^{-1}, \\
	f_{\rm K}^U &= \left( e^{\beta_{\rm K} (\varepsilon + U)} + 1 \right)^{-1},
\end{align*}
with $\beta_{\rm K} = 1 / (k_{\rm B} T_{\rm K})$ the inverse temperature of reservoir $K$.

This system of coupled differential equations admits analytical solutions within each stroke of the Otto cycle. Using these solutions, we compute the thermodynamic quantities of interest via Eqs.~\eqref{heat_c}--\eqref{worktot} employing the weak-coupling version, where the the effective Hamiltonian $K_{\rm S}$ is replaced by the bare system Hamiltonian $H_{\rm S}$. In particular, the heat exchanged with the hot and cold reservoirs, as well as the total work output per cycle, are given by 
\begin{align*}
	Q_{\rm h} &= \varepsilon_{\rm h} \Delta \rho_\uparrow + (2\varepsilon_{\rm h} + U)\, \Delta \rho_{\uparrow\downarrow}, \\
	Q_{\rm c} &= -\varepsilon_{\rm c} \Delta \rho_\uparrow - (2\varepsilon_{\rm c} + U)\, \Delta \rho_{\uparrow\downarrow}, \\
	W_{\text{tot}} &= -(\varepsilon_{\rm h} - \varepsilon_{\rm c})(\Delta \rho_\uparrow + 2\Delta \rho_{\uparrow\downarrow}),
\end{align*}
where the changes in population during the hot and cold strokes are defined as 
\begin{align*}
	\Delta \rho_\uparrow &= \rho_\uparrow(\mathcal{T}/3) - \rho_\uparrow(5\mathcal{T}/6), \\
	\Delta \rho_{\uparrow\downarrow} &= \rho_{\uparrow\downarrow}(\mathcal{T}/3) - \rho_{\uparrow\downarrow}(5\mathcal{T}/6).
\end{align*}

In the non-interacting case ($U=0$), the phase diagram of the Otto cycle in the $(\varepsilon_{\rm h}, \varepsilon_{\rm c})$ plane exhibits a symmetric structure centered at the origin, as shown in Fig.~\ref{fig:otto_working_regimes} (top left panel). In this case, the system operates as a heat engine when both energy levels have the same sign and $  |\varepsilon_{\rm h}| T_{\rm c}/T_{\rm h} <|\varepsilon_{\rm c}| < |\varepsilon_{\rm h}|$, converting heat from the hot reservoir into useful work. When $  0 <|\varepsilon_{\rm c}| < |\varepsilon_{\rm h}| T_{\rm c}/T_{\rm h}$, the system enters the refrigeration regime, in which external work is used to extract heat from the cold reservoir. The exact position and shape of these regimes depend sensitively on the temperature ratio $T_{\rm c} / T_{\rm h}$.

Upon introducing a finite Coulomb interaction ($U \neq 0$), the center of symmetry of the diagram shifts to $(\varepsilon_{\rm h}, \varepsilon_{\rm c}) = (-U/2, -U/2)$ due to the energetic elevation of the doubly occupied state, as visible in the middle and right panels of Fig.~\ref{fig:otto_working_regimes}. As $U$ increases, the refrigeration region progressively shrinks and eventually disappears for large $U$, being replaced by dissipative regimes (accelerator or heater), as heat absorption from the cold reservoir becomes energetically suppressed.
This behavior originates from the fact that a large Coulomb interaction strongly suppresses fluctuations into the doubly occupied state, effectively blocking the transitions that would otherwise allow the system to extract heat from the cold reservoir.

For the heat-engine regime, the efficiency of the Otto cycle in the non-interacting case ($U=0$) reduces to the familiar expression,
\[
\eta_0 = 1 - \frac{\varepsilon_{\rm c}}{\varepsilon_{\rm h}}.
\]

For $U \neq 0$, using the weak-coupling expressions for work and heat, the efficiency can be written as
\begin{align*}
	\eta_{\rm U} 
	&= \frac{(\varepsilon_{\rm h} - \varepsilon_{\rm c})(\Delta \rho_\uparrow + 2\Delta \rho_{\uparrow\downarrow})}{\varepsilon_{\rm h} \Delta \rho_\uparrow + (2\varepsilon_{\rm h} + U)\, \Delta \rho_{\uparrow\downarrow}} \\
	&= \eta_0\frac{
		1 + 2\,\dfrac{\Delta\rho_{\uparrow\downarrow}}{\Delta\rho_\uparrow}
	}{
		1 + 
		\left(2 + \dfrac{U}{\varepsilon_{\rm h}}\right)
		\dfrac{\Delta\rho_{\uparrow\downarrow}}{\Delta\rho_\uparrow}
	}.
\end{align*}

The conditions for $\eta_{\rm U}$ to surpass $\eta_0$ can be easily obtained,
\begin{align*}
	\eta_{\rm U} > \eta_0 \quad \Longleftrightarrow \quad
	\begin{cases}
		2 + \dfrac{U}{\varepsilon_{\rm h}} < -\,\dfrac{\Delta \rho_\uparrow}{\Delta \rho_{\uparrow\downarrow}}, & \varepsilon_{\rm h} > 0 ,\\[1em]
		2 + \dfrac{U}{\varepsilon_{\rm h}} > -\,\dfrac{\Delta \rho_\uparrow}{\Delta \rho_{\uparrow\downarrow}}, & \varepsilon_{\rm h} < 0 .
	\end{cases}
\end{align*}

In the considered parameter regime, the first condition is never satisfied. In contrast, for $\varepsilon_{\rm h} < 0$, the second condition can be fulfilled. Interestingly, this occurs only when
\[
\frac{\Delta\rho_{\uparrow\downarrow}}{\Delta\rho_\uparrow} < 0.
\]
Here, $\Delta\rho_\uparrow > 0$ holds, as expected physically: in a heat engine, the population of the singly occupied state naturally increases during the hot stroke. Therefore, the condition above reduces to 
\[
\Delta\rho_{\uparrow\downarrow} < 0 ,
\]
that is, if the occupation of the doubly occupied state decreases during the hot stroke of the cycle.
Physically, this corresponds to a situation in which the system empties the high-energy doubly occupied state during the hot phase and refills it during the cold phase, effectively reducing the heat input required from the hot reservoir to produce the same output of work.

A phase-space analysis of the sign of $\Delta \rho_{\uparrow\downarrow}$ reveals that this mechanism occurs predominantly when both energy levels $\varepsilon_{\rm h,c}$ lie below the Fermi level, i.e.
\[
2\varepsilon_{\rm h,c} + U < 0 .
\]

\noindent
In this regime, the asymmetric occupation of the interaction–shifted and lifetime–broadened energy levels becomes favorable, highlighting how Coulomb interaction can be exploited to enhance thermodynamic efficiency even in the weak-coupling limit.

\bibliographystyle{apsrev4-2}
\bibliography{biblio_nonmarkov}

\end{document}